\documentclass[fleqn]{aa}  
\usepackage{graphicx}
\usepackage{amsmath}
\usepackage{comment}
\usepackage{txfonts}
\usepackage[normalem]{ulem}
\usepackage{hyperref}
\hypersetup{
	colorlinks=true,
    linkcolor=blue,
    filecolor=blue,      
    urlcolor=blue,
    citecolor=blue,
}

\begin{document} 

\newcommand{\ngc}{NGC~1068}
\newcommand{\ic}{IC~4329A}

   \title{Nonthermal processes in hot accretion flows onto supermassive black holes: An inhomogeneous  model}
   \author{
          E. M. Guti\'errez \inst{1}
          \and
          F. L. Vieyro \inst{1,2}
          \and
          G. E. Romero \inst{1,2}
    }

   \institute{Instituto Argentino de Radioastronom\'ia (IAR, CONICET/CIC/UNLP), C.C.5, (1894) Villa Elisa, Buenos Aires, Argentina
              \email{emgutierrez@iar.unlp.edu.ar}
         \and
             Facultad de Ciencias Astron\'omicas y Geof\'isicas, Universidad Nacional de La Plata, Paseo del Bosque s/n, 1900 La Plata, Buenos Aires, Argentina
    }

   \date{Submitted}

\titlerunning{Nonthermal processes around supermassive black holes}
\authorrunning{Guti\'errez, Vieyro \& Romero}

  \abstract
     {Many low-redshift active galactic nuclei harbor a supermassive black hole accreting matter at low or medium rates. At such rates, the accretion flow usually consists of a cold optically thick disk, plus a hot, low density, collisionless corona. In the latter component, charged particles can be accelerated to high energies by various mechanisms.}
   {We aim to investigate, in detail, nonthermal processes in hot accretion flows onto supermassive black holes, covering a wide range of accretion rates and luminosities.}
  {We developed a model consisting of a thin Shakura-Sunyaev disk plus an inner hot accretion flow or corona, modeled as a radiatively inefficient accretion flow, where nonthermal processes take place. We solved the transport equations for relativistic particles
  and estimated the spectral energy distributions resulting from nonthermal interactions between the various particle species and the fields in the source.}
   {We covered a variety of scenarios, from low accretion rates up to 10\% of the Eddington limit, and identified the relevant cooling mechanisms in each case. The presence of hadrons in the hot flow is decisive for the spectral shape, giving rise to secondary particles and gamma-ray cascades. We applied our model to the source IC 4329A, confirming earlier results which showed evidence of nonthermal particles in the corona.}
   {}

\keywords{Relativistic processes -- radiation mechanisms: nonthermal -- black hole physics -- accretion, accretion disks -- galaxies: active}

\maketitle
%

\section{Introduction}

The accretion of matter and magnetic fields onto compact objects is the mechanism responsible for powering the most energetic phenomena known in nature. In particular, active galactic nuclei (AGNs), which consist of a supermassive black hole accreting material from the central region of a host galaxy, are the sources that dominate the gamma-ray sky \citep{Abdollahi2020}. Most gamma-ray emitting AGNs are blazars, whose luminosity is beamed by the relativistic bulk motion of the plasma in a jet. Some extragalactic gamma-ray sources, however, are nonblazars: radio galaxies, usually associated with low-luminosity AGNs (LLAGNs), Narrow Line Seyfert 1 galaxies, which accrete close to the Eddington limit \citep{Rieger2017}, and a few Seyfert 2 galaxies \citep{wojaczynski2015,fermi2019}. The latter usually present both an AGN and a starburst, and it is still unclear what is the relative contribution of each component to the total gamma-ray emission \citep{wojaczynski2015}.

From a physical point of view, accretion flows are classified into different regimes depending mainly on the accretion rate (\citealt{chen1995}, see also \citealt{begelman2014}). At low accretion rates, the flow behaves as an optically thin radiatively inefficient accretion flow (RIAF). In many situations of interest, the RIAF may coexist with a standard Shakura-Sunyaev disk (SSD,  \citealt{shakura1973,novikov1973}). This scenario leads to the so-called SSD+RIAF model \citep{Bisnovatyi1977,narayan1996,dove1997}, which has been applied to explain the various spectral states of black hole binaries (BHBs, \citealt{narayan1996,poutanen1997,esin1997,esin1998}), the broadband spectrum of LLAGNs \citep{maraschi2003,nemmen2014}, and some Seyfert galaxies (\citealt{chiang2003,yuan2004b,yuan2014} and references therein). This family of models considers that an outer cold thin disk is truncated at a radius $R_{\rm tr} > R_{\rm ISCO}$, where $R_{\rm ISCO}$ is the radius of the innermost stable circular orbit (ISCO). At the truncation radius, the disk evaporates into an inner hot advection-dominated accretion flow that extends down to the black hole event horizon. Independently of the details of the physical mechanisms for this transition (see, e.g., \citealt{abramowicz1988}), there must be a region of overlap between the two states. Moreover, an SSD extending down to the ISCO is insufficient to explain many features of the observed X-ray phenomenology in many luminous AGNs, and the presence of a hot optically thin plasma above and below the disk is also required \citep{Bisnovatyi1977,poutanen1998}.

Unlike SSDs, which are expected to be dense enough for the plasma to thermalize quickly, hot accretion flows (HAFs\footnote{ Throughout the work, we use the terms HAF, RIAF, ``corona'', and ``advection-dominated accretion flow'' to refer to the same physical system, namely the hot, inflated, optically thin component of the accreting structure. See \citet{yuan2014} for a complete review of these flows.} ) can be weakly collisional or even collisionless and thus suitable for the occurrence of particle acceleration and nonthermal processes \citep{mahadevan1997b}. Although the gamma-ray emission in AGNs is usually associated with the presence of a relativistic jet, the detection of high-energy radiation from radio-quiet AGNs motivates the investigation of the contribution of nonthermal processes in accretion flows to the overall energetic output of these sources. Moreover, the gamma-ray radiation from a few Seyfert 2 plus starburst galaxies not only shows variability but also seems to lie well above both the known IR/$\gamma$ and radio/$\gamma$ correlation; this suggests that putative AGNs might be powering most of their gamma-ray output \citep{wojaczynski2017,peng2019}.

There is also significant observational evidence that supports the idea that particles are accelerated in HAFs: The steady radio emission from Sgr A* is thought to be produced by a population of nonthermal electrons within the HAF that feeds the central black hole \citep{yuan2003,liu2013}. Additionally, the multiwavelength flaring activity of this source is likely related to nonthermal activity in the flow \citep{yuan2003,yuan2004a,gutierrez2020a,dexter2020}. Recently, \citet{inoue2018} found evidence of nonthermal electron activity occurring in the corona of more luminous AGNs such as Seyfert I galaxies.
The presence of nonthermal protons in HAFs is more difficult to trace directly, but it can be indirectly inferred from the detection of neutrinos or cosmic rays, as well as gamma-ray cascade emission \citep{kimura2015,kimura2019,inoue2020}. From a theoretical point of view, protons thermalize much slower than electrons, so it is expected that they retain a longer memory of the heating or acceleration process. Moreover, recent Particle-In-Cell (PIC) simulations show that protons are accelerated much more efficiently than electrons in turbulent environments such as accretion flows \citep{zhdankin2019}.

The study of the transport of nonthermal particles in coronae or HAFs around supermassive and stellar-mass black holes is a highly complex field. It involves numerous physical ingredients: the structure and dynamics of a disk, the thermal background --including the thermal coupling of electrons and protons and various mechanisms of cooling and heating of the gas--, and the possible presence of nonthermal particles with their corresponding acceleration and transport. Several aspects of the problem have been previously explored in the literature with different approaches: Some studies include only leptonic contributions (e.g., \citealt{coppi1992,vurm2009,veledina2011,Bandyopadhyay2019}), whereas other works include hadronic processes \citep{romero2010,vieyro2012,rodriguezramirez2019,inoue2019,kimura2015,kimura2019}. Many previous models adopted a homogeneous spherical corona and rely on the one-zone approximation (e.g., \citealt{vurm2009, romero2010, vieyro2012, kimura2015, kimura2019, inoue2019}). On the other hand, those models using actual hydrodynamic solutions for the RIAF do not usually solve the nonthermal transport equations but assume or fit the energy distribution for the relativistic particles (e.g., \citealt{ozel2000, yuan2003, wojaczynski2015,Bandyopadhyay2019}). Besides, different techniques are also used to address the problem: Detailed numerical simulations allow treating in much more detail nonlinear phenomena, and are essential when one wants to tackle time evolution or multidimensional phenomena (such as outflows)  (e.g., \citealt{hilburn2010,yuan2014,chael2017}). Semi-analytical procedures, on the other hand, allow to study most of the relevant physical processes that take place globally within the flow, with the advantage that are more versatile to give a physical grasp of the situation \citep[e.g.,][]{vieyro2012,kimura2019}.

In this work, we develop a new model of an HAF onto a black hole with focus on the nonthermal processes. We use a semi-analytical treatment that combines several of the ingredients mentioned above: a disk with both hot and cold components, thermal and nonthermal particles, and their interactions and transport. In a first step, we solve the hydrodynamic equations to obtain the accretion flow structure for various accretion regimes. Then, we calculate the electromagnetic emission of the thermal electrons in the flow. Once the thermal background is set, we inject a population of relativistic particles, both electrons and protons and estimate the outputs (photons and secondary particles) resulting from their interactions with the environment. To this end, we solve the transport equation for each particle species, and compute the spectral energy distribution (SED) produced by all relevant processes, taking into account the radial dependence in the physical properties of the flow.

The remainder of this article is organized as follows: In Section \ref{sec:accretion_model}, we present in detail the accretion model. In Section \ref{sec:nonthermal}, we discuss the acceleration and transport of the relativistic particles in the HAF and how we treat them. In Section \ref{sec:seds} we describe the calculation of the SED resulting from all the nonthermal processes discussed in the previous sections, for a set of specific models. In Section \ref{sec:results} we present the general results, and in Section \ref{sec:IC4329A} we apply the model to the Seyfert galaxy \ic. In Section \ref{sec:discussion} we discuss various phenomena where nonthermal processes in HAFs might play important roles. Finally, we present a summary and our conclusions in Section \ref{sec:conclusions}.

\section{Accretion flow model}
\label{sec:accretion_model}

\begin{figure}
    \centering
    \includegraphics[width=\linewidth]{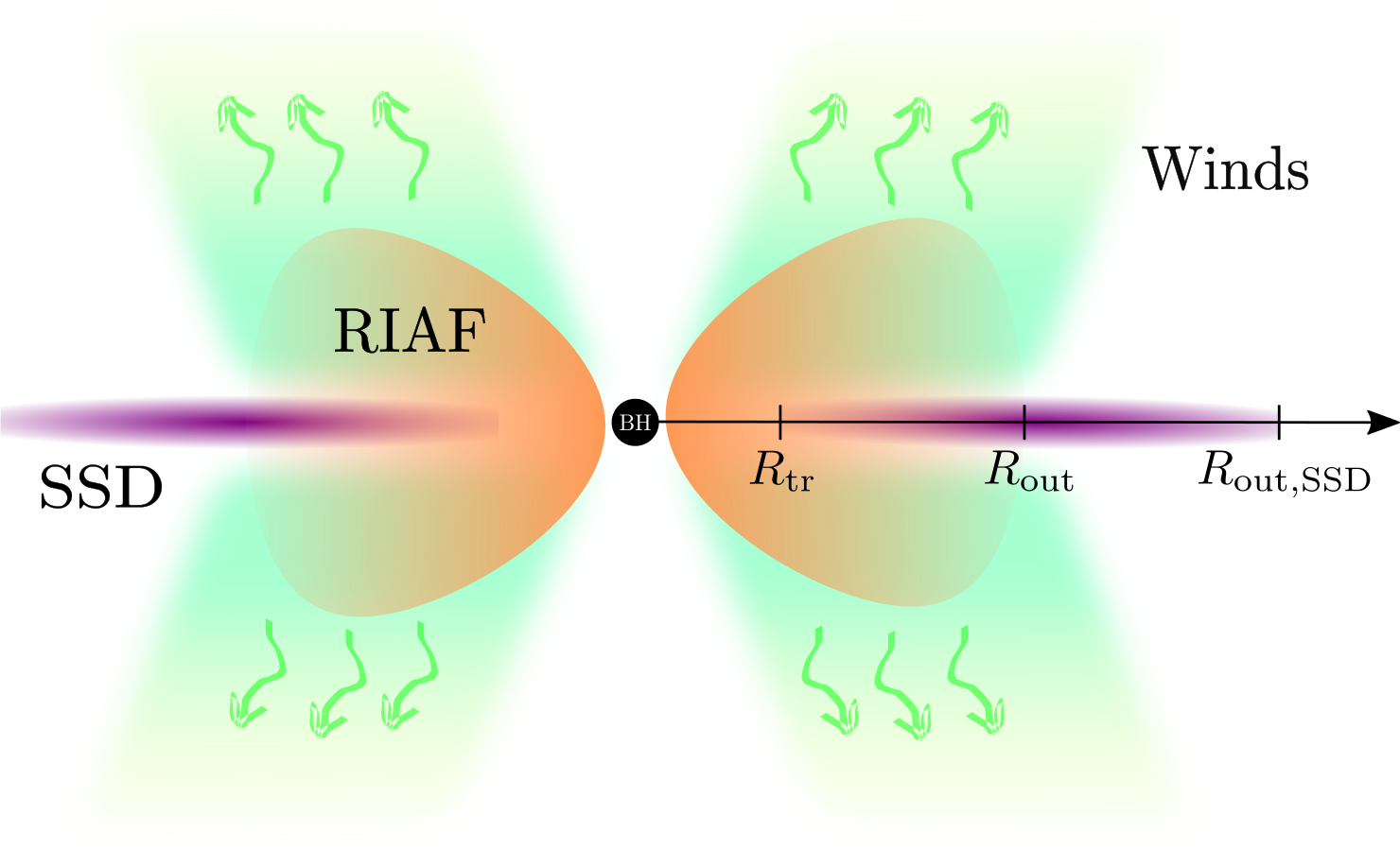}
    \caption{Cartoon representing the accretion system. A jet might be present but it is not considered in our model.}
    \label{fig:cartoon}
\end{figure}

The model is constructed so that it is general enough to be applied to a wide variety of accretion flows onto black holes, from those in LLAGNs like Sgr A*, which are usually modeled as pure RIAFs \citep{narayan1998,yuan2003}, to those powering moderately luminous Seyfert galaxies. In the latter class of objects, an SSD disk may penetrate down to the ISCO \citep{wojaczynski2015,inoue2019}. A general picture with the basic geometry of the model is shown in Fig. \ref{fig:cartoon}. It consists of an SSD extending from a radius $R_{\rm out,SSD}\sim 10^{5-6}R_{\rm S}$ down to the truncation radius $R_{\rm tr}$, and an HAF extending from a radius $R_{\rm out}\sim 10^{2-4}R_{\rm S} \geq R_{\rm tr}$ down to the event horizon. Here, $R_{\rm S}=2GM_{\rm BH}/c^2$ denotes the Schwarzschild radius, $M_{\rm BH}$ is the black hole mass, $G$ is the gravitational constant, and $c$ is the speed of light in vacuum. In the region $R_{\rm tr}<R<R_{\rm out}$ the two states coexist, and the HAF plays the role of a corona above and below the thin disk. The total accretion rate is divided into the SSD and the corona:
\begin{equation}
    \dot{M}(R) = \dot{M}_{\rm d}(R)+\dot{M}_{\rm c}(R),
\end{equation}
where $\dot{M}(R)$ is the total accretion rate at radius $R$, $\dot{M}_{\rm c}(R)$ is the accretion rate through the corona, and $\dot{M}_{\rm d}(R)$ is the accretion rate through the SSD. We write
\begin{flalign}
\dot{M}(R) &= \dot{M}_{\rm out} \left[ 1-w(R) \right], \\
    \dot{M}_{\rm d}(R) &= \dot{M}_{\rm out}~ f(R), \\
    \dot{M}_{\rm c}(R) &= \dot{M}_{\rm out}~ g(R),
\end{flalign}
so that $f(R)+g(R)=1-w(R)$, where $w(R)$ accounts for the total mass loss rate via winds integrated from $R_{\rm out}$ to $R$.
Here, $\dot{M}_{\rm out}$ is the accretion rate at the outer boundary of the system, and $f(R)$ makes the transition between the two components smooth. It is well-known that HAFs present magneto-centrifugal winds that decrease the amount of matter that actually reaches the black hole \citep{narayan1995,stone1999,yuan2012}. This decrease is usually parameterized in a phenomenological way as \citep{blandford1999}
\begin{equation}
    \dot{M}_{\rm c} (R) = \dot{M}_{\rm out} \left( \frac{R}{R_{\rm out}} \right)^s,
\end{equation}
with $0<s<1$ (see \citealt{yuan2014}, Sect. 3.4, for a discussion on the validity of this approximation).
Hence, we define $g(R)$ in such a way that when $f(R)=0$, that is to say, when the SSD has completely evaporated, $\dot{M}_{\rm c} \propto R^s$. Expressions for $f(R)$, $g(R)$ and $w(R)$ are given in Appendix \ref{ap:acc_rates}.%

\subsection{Hot accretion flow}

We modeled the hot, inflated, optically thin inner component of the accretion flow as an RIAF. Since this is the site where we aim to explore nonthermal phenomena, we treat with detail both the hydrodynamics and the radiative outputs of this component.

\subsubsection{Hydrodynamics}

We obtain the hydrodynamical structure of the RIAF (mass density $\rho$, magnetic field $B$, proton and electron temperatures $T_{\rm p}$, $T_{\rm e}$, etc) by solving the height-integrated, steady-state hydrodynamic equations \citep{abramowicz1988,yuan2003} via the shooting method with appropriate boundary conditions\footnote{When $R_{\rm out} \sim 10^3R_{\rm S}$, the outer proton temperature is $T_{\rm out,p}=0.2T_{\rm vir}$, and the electron temperature is $T_{\rm out,e}=0.19T_{\rm vir}$; here, $T_{\rm vir} = 3.6 \times 10^{12}(R_{\rm S}/R)~{\rm K}$ is the virial temperature. The Mach number at $R_{\rm out}$ is $M_{\rm s} \equiv v/c_{\rm s} = 0.2$.} (\citealt{yuan2000}). The remaining parameters to determine the structure of the flow are the viscosity $\alpha$-parameter, the plasma $\beta$-parameter (gas pressure to magnetic pressure ratio), and the fraction of energy released by turbulence that directly heats electrons, $\delta$. Early studies on ADAFs considered $\delta$ to be small ($\lesssim 10^{-3}$), but more recent numerical \citep{quataert1999,sharma2007} and observational works \citep{yu2011} suggest that it may be as high as $\sim 0.5$. Unless explicitly stated, we fix these parameters to standard values: $\alpha=0.1$, $\beta=9$, $\delta=0.1$ and $s=0.3$ \citep{yuan2014}.

To mimic Schwarzschild spacetime, we consider a Paczynski-Wiita gravitational potential \citep{paczynsky1980} with a small additional relativistic correction in the radial velocity \citep{yuan2006}. Since the accreted gas reaches the black hole event horizon at the speed of light, the hydrodynamical solution must be transonic. The calculation necessary to obtain such a transonic solution involves adjusting iteratively an eigenvalue (the specific angular momentum accreted by the black hole) in such a way that the sonic radius is crossed smoothly (see \citealt{yuan2000} for details).

Although we consider that the plasma has both thermal and nonthermal components, we assume that the thermal gas dominates energetically. Then, the kinetic energy density of protons $({\rm p})$ and electrons $({\rm e})$ is simply \citep{chandrasekhar1939}
\begin{equation}
	u_{\rm q} \approx u_{\rm q}^{\rm th} = a(\theta_{\rm q}) n_{\rm q} m_{\rm q} c^2 \theta_{\rm q},
\end{equation}
where
\begin{equation}
	a(\theta_{\rm q}) = \frac{1}{\theta_{\rm q}} \left[ \frac{3 K_3(1/\theta_{\rm q})+K_1(1/\theta_{\rm q})}
	{4K_2(1/\theta_{\rm q})}-1 \right].
\end{equation}
Here, $K_n$ are the modified Bessel functions of $n$th order, $\theta_{\rm q} = k_{\rm B}T_{\rm q}/m_{\rm q} c^2$, $m_{\rm q}$ and $n_{\rm q}$ are the mass and the number density of particles of the species ${\rm q=p,e}$; $k_{\rm B}$ is the Boltzmann constant.

\subsubsection{Thermal emission}

Radiatively inefficient accretion flows are optically thin and hence both the cooling function and the shape of the emitted spectrum depend strongly on the details of the radiative processes that take place in the flow. Electrons reach relativistic temperatures ($\theta_{\rm e} \gtrsim 1$) and cool via synchrotron radiation, Bremssthralung radiation, and inverse Compton up-scattering of low-energy photons. Whereas the first two processes are local, inverse Compton is not: Photons may suffer multiple scatterings in different regions of the flow before escaping.
Nevertheless, to estimate the cooling function of the flow, a local treatment is usually adopted giving fairly accurate results \citep{dermer1991,esin1996,manmoto1997}. We use this approximation when solving the hydrodynamical structure, though we do take into account nonlocal scatterings in the calculation of the spectrum (see Sect. \ref{sec:seds}).

For thermal protons, the three cooling mechanisms mentioned above are completely negligible, but since they reach much higher temperatures than electrons (almost virial, $\sim 10^{12}~{\rm K}$ in the inner regions), those at the tail of the Maxwellian distribution have sufficient energy to produce neutral pions (and thus gamma rays) via proton-proton (${\rm pp}$) collisions\footnote{Neutral pions decay with a mean lifetime of $8.4\times 10^{-17}~{\rm s}$ into two gamma rays $\pi^0 \rightarrow 2\gamma$.} \citep{mahadevan1997a,oka2003}.

We solve the hydrodynamic equations and obtain the temperatures $T_{\rm e,p}(R)$, magnetic field $B(R)$, mass density $\rho(R)$, radial velocity $v(R)$, and height $H(R)$ as a function of the distance from the event horizon. To calculate the thermal SED emitted by the flow, we divide its volume into $N$ logarithmically-spaced cylindrical shells, and solve the radiative transfer taking into account the coupling (through Comptonization) among them. When an SSD is present in the model, we consider its emission as a seed for Compton scattering as well. In Section \ref{sec:seds}, we describe in detail the radiation transfer calculation.


\subsection{Thin disk} \label{subsec:thin_disk}

We calculate the emission from the SSD using the standard multicolor blackbody method (e.g., \citealt{frank2002}). We take into account the coupling between the SSD and the RIAF by means of two mechanisms: heating of the SSD by absorption of incident radiation from the RIAF and inverse Compton scattering of SSD photons by hot electrons in the RIAF \citep{narayan1997}. For simplicity, we ignore X-ray reflection in the SSD; it should be taking into account, however, to investigate in detail the X-ray phenomenology of the AGN.

\subsection{Model parameters}
\label{sec:model_parameters}

\begin{figure}
    \centering
    \includegraphics[width=0.9\linewidth]{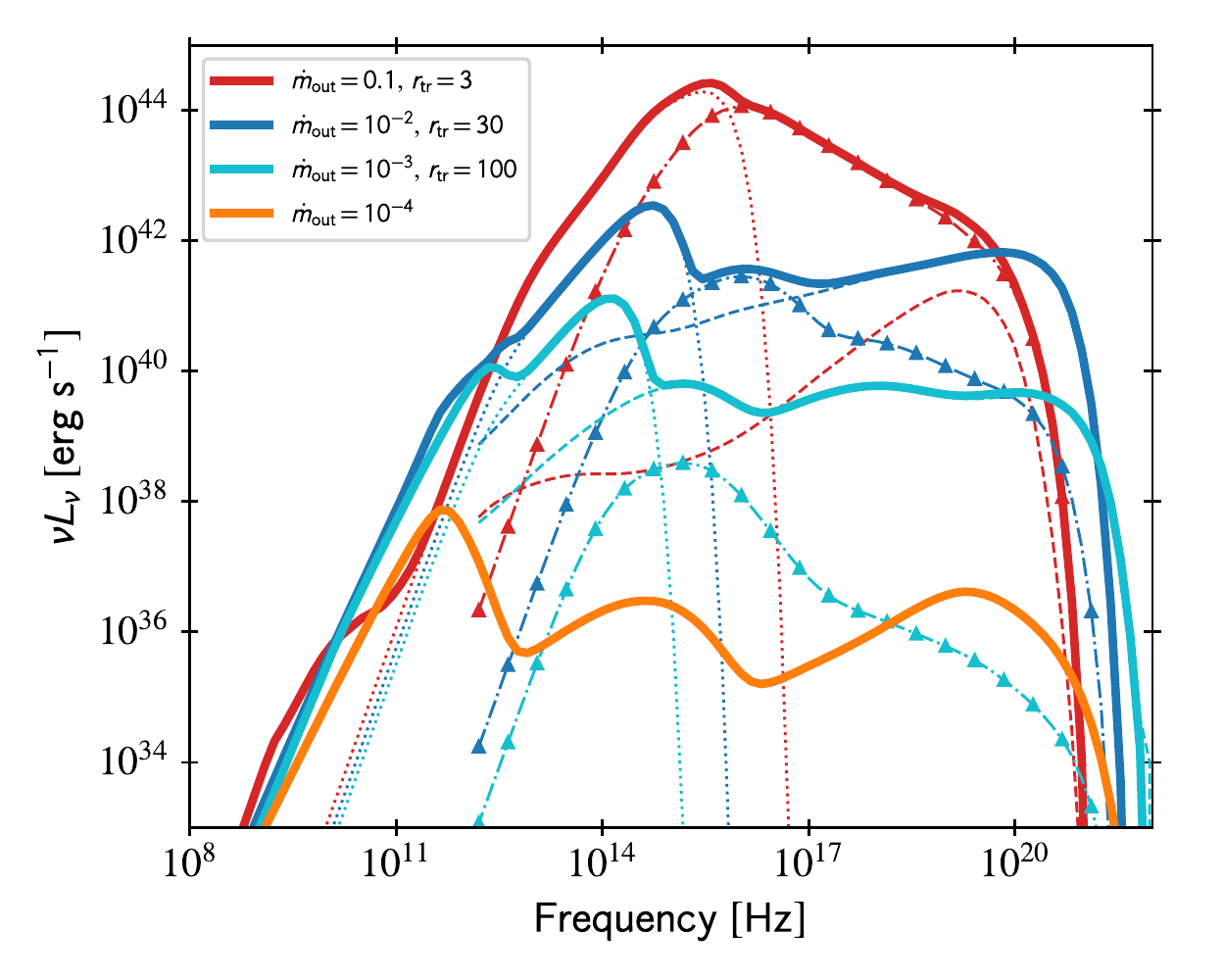}
    \caption{ Thermal emission from the accretion flow around a supermassive black hole of mass $M_{\rm BH}=10^8 M_\odot$ for four different models. In the three flows with highest accretion rate, we show in thin dotted lines the direct emission from the thin disk, in dot-dashed lines with triangle markers the Comptonization of the thin-disk photons, and in dashed lines the Comptonization of the local Synchrotron and Bremsstrahlung photons from the RIAF.}
    \label{fig:thermal_emission}
\end{figure}

From here on, we shall express the black hole mass in units of solar masses: $M_{\rm BH} = m~M_\odot$, and accretion rates in units of the Eddington accretion rate,
$\dot{M}=\dot{m}~\dot{M}_{\rm Edd}$,
where $\dot{M}_{\rm Edd}=L_{\rm Edd}/\eta c^2$, $L_{\rm Edd}=1.26\times 10^{38}~m$ erg s$^{-1}$ is the Eddington luminosity, and $\eta=0.1$ is the radiative efficiency\footnote{The actual radiative efficiency is in general much lower than $0.1$.}. We also express radii in units of the Schwarzschild radius: $R = r~R_{\rm S}$.
Of the many parameters of the model, the accretion rate is the most important one. Hence, we fix the mass of the central black hole to $m=10^8$, and the HAF outer radius to $r_{\rm out}=1000$, and we vary the outer accretion rate. We explore four different scenarios with increasing accretion rate: $\dot{m}_{\rm out}=10^{-4},~10^{-3},~10^{-2},~10^{-1}$. In the latter three scenarios, we assume that a thin disk penetrates down to $r_{\rm tr}=100$, $30$, and $3$\footnote{Accretion flows with very low accretion rate ($\dot{m}_{\rm out} \lesssim 10^{-4}$) are usually modeled as pure RIAFs. Although an optically thick component might be present at outer radii, its effect would be negligible.}. We also make the transition softer as the disk penetrates more into the RIAFs, thus we take $b=2,1,0.5$, respectively (see Appendix \ref{ap:acc_rates}). This is reasonable since a high value of $b$ implies an abrupt transition between the two accretion states at the truncation radius, and hence a too small corona when the SSD penetrates down to the innermost regions.
Figure \ref{fig:thermal_emission} shows the SED resulting from thermal processes in each of these four scenarios. At low accretion rates, three peaks can be clearly distinguished in the spectrum, and they correspond to synchrotron emission, a first inverse Compton scattering of these synchrotron photons, and Bremsstrahlung radiation, respectively. As the accretion rate increases, the SSD penetrates deeper into the HAF and its blackbody emission starts to compete with the synchrotron emission from the HAF, and second and higher-order inverse Compton scatterings overtake the Bremsstrahlung X-ray radiation, hardening the spectrum. Also, the Comptonization of the SSD photons increases. In the most luminous scenario, the emission is completely dominated by the radiation from an SSD penetrating down to the ISCO, and the power-law tail is caused by the superposition of many inverse Compton scatterings of the blackbody photons from the SSD by the hot electrons in the corona. The spectral index\footnote{ The spectral index of the power-law is consistent with the analytical expression $\alpha=-\log \tau_{\rm T}/\log A$, where $\tau_{\rm T}$ is the Thomson optical depth and $A=1+4\theta_{\rm e} + 16\theta_{\rm e}^2$ \citep{rybicki1979}} of the spectrum in this latter scenario is determined mainly by the $b$ parameter (see Figure \ref{fig:IC4329A} for comparison with this case).

\section{Nonthermal particles}
\label{sec:nonthermal}

Many numerical studies show that particles can be accelerated in hot accretion flows (see, e.g., \citealt{li1997,lynn2014}).
This is expected since an RIAF is a collisionless plasma with strong and turbulent magnetic fields. In such an environment, several mechanisms can accelerate particles to relativistic energies. The most plausible ones are magnetic reconnection \citep{hoshino2012,degouveiadalpino2010}, stochastic acceleration by turbulence \citep{dermer1996,zhdankin2019}, and diffusive acceleration mediated by shocks\footnote{Only in regions where the magnetic energy density is much smaller than the gas energy density.} \citep{drury1983,blandford1987}.

\subsection{Particle acceleration and transport}
\label{subsec:acceleration}

Regardless of the acceleration mechanism, we assume that a fraction of the particles is pushed out from the thermal distribution and get accelerated to high energies. We assume that after being accelerated, the particle population can be described by a power-law injection function of the form

\begin{equation}
    Q(\gamma,r) = Q_0(r) ~ \gamma^{-p} \exp \left[-\gamma/\gamma_{\rm cut}(r)\right],
\end{equation}
characteristic of diffusive acceleration mechanisms. We shall consider two values for the spectral index: $p = 2$, where the power is distributed equally along the whole energy range, and $p=1.2$, where most of the power is in the highest energy particles. The cutoff Lorentz factor at each radius is estimated by the balance between the acceleration timescale and the cooling/escape timescale (see Section \ref{subsec:timescales}). The transport equation that governs the evolution of the population of nonthermal particles is
\begin{equation}
\label{eq:transport_eq}
    \frac{\partial N}{\partial t} + \vec{\nabla} \cdot \left( \vec{v} N - D_r \vec{\nabla}N \right) = \frac{\partial}{\partial \gamma} \Big ( |\dot{\gamma}_{\rm rad}| N \Big )
    - \frac{N}{t_{\rm esc}} + Q,
\end{equation}
where $N(\gamma,r)d\gamma$ is the number of particles per unit volume whose Lorentz factor lies in the range $(\gamma,\gamma+d\gamma)$, $D_r$ is the radial diffusion coefficient, $\dot{\gamma}_{\rm rad}$ is the radiative cooling rate, and $t_{\rm esc}$ is the escape timescale\footnote{We assume escape in the vertical direction.}. The spatial derivatives on the left-hand-side in Eq. \ref{eq:transport_eq} are responsible for the radial transport: The first term represents advection toward the black hole, and the second term represents radial diffusion.
In cylindrical coordinates, assuming axisymmetry, homogeneity in the vertical direction, and steady state, we can re-write Eq. \ref{eq:transport_eq} in the form
\begin{equation}
\label{eq:transport_DSA}
    \frac{\partial}{\partial r} \left[ \mathcal{A}(\gamma,r) \frac{\partial \tilde{N}}{\partial r} + \mathcal{B}(\gamma,r) \tilde{N}\right] - \frac{\partial}{\partial \gamma} \Big ( |\dot{\gamma}_{\rm rad}| \tilde{N} \Big )
    + \frac{\tilde{N}}{t_{\rm esc}} = \tilde{Q},
\end{equation}
where $\tilde{N}(\gamma,r):= r N(\gamma,r)$, $\mathcal{A}(\gamma,r):=-D_r(\gamma,r)$, and $\mathcal{B}(\gamma,r):=-\mathcal{A}/r+v_r(r)$.

Equation \ref{eq:transport_DSA} is a two-dimensional advection-diffusion equation in the $(\gamma,r)$-space. Nevertheless, since energy and radial transport have in general very different natural timescales, Eq. \ref{eq:transport_DSA} can be simplified when the transport in one dimension dominates over the other. For electrons, radiative cooling is much faster than radial transport at almost all energies. Only electrons with $\gamma < 10$ in the innermost regions can be advected without being completely cooled (see Section \ref{subsec:timescales}), and we can neglect the first term and solve the equation under the one-zone approximation at each cylindrical shell.
The opposite is true for protons, meaning that they carry their energy throughout the flow via advection and diffusion. Radiative cooling, however, cannot be neglected at the highest energies. To tackle this effect, we approximate the second term in the left-hand side of Eq. \ref{eq:transport_DSA} as $\approx - \tilde{N} / t_{\rm rad}$, where $t_{\rm rad} := |\dot{\gamma}_{\rm rad}|/\gamma$ is the cooling timescale.
We solve Eq. \ref{eq:transport_DSA} via finite differences with the \citet{chang-cooper1970} discretization as outlined in \citet{park1996}. 

We normalize the injection function by assuming that a fraction $\varepsilon_{\rm NT}$ of the total accretion power goes to nonthermal particles. This power is divided into protons and electrons with the prescription $L_{\rm e} = \varepsilon_{\rm e} L_{\rm NT}$, and $L_{\rm p} = \varepsilon_{\rm p} L_{\rm NT} = (1-\varepsilon_{\rm e})L_{\rm NT}$. In addition, we must give a functional form for the radial dependence of the injected nonthermal power. We conservatively assume $Q_0(r) \propto B(r)~ u_{\rm th}(r)$. Thus,
\begin{multline}
	L_{\rm e,p} = \varepsilon_{\rm NT}~ \varepsilon_{\rm e,p}~ \dot{M}_{\rm out}c^2= \\ q_0~\int dV~ B(r)~u_{\rm th}(r) \int_{\gamma_{\rm min}}^{\gamma_{\rm max}} d\gamma~\gamma^{-p+1},
\end{multline}
where $q_0$ is a normalization constant.
Finally, after solving the transport equation we check that $p_{\rm CR}\footnote{The cosmic ray pressure at the position $r$ is $p_{\rm CR}(r):= \frac{1}{3}m_{\rm p} c^2\int d\gamma ~\gamma ~N(r,\gamma)$.} \lesssim 0.2p_{\rm gas}$ for all radii, assuring that our initial assumption that nonthermal particles do not affect the flow structure is fulfilled.

\subsection{Timescales and radiation processes}
\label{subsec:timescales}

The timescale for the acceleration of a particle of mass $m$ and charge $e$ to an energy $\gamma mc^2$ depends on the acceleration mechanism.
We enclose the uncertainties into the acceleration efficiency parameter $\eta_{\rm acc} < 1$, and write the acceleration timescale as (e.g., \citealt{aharonian2002})
\begin{equation}
\label{eq:acc_timescale}
	t_{\rm acc}(\gamma) \sim \eta_{\rm acc}^{-1}\frac{r_{\rm L}}{c} = \eta_{\rm acc}^{-1}\frac{\gamma m c}{ e B },
\end{equation}
where $r_{\rm L}$ is the relativistic Larmor radius of the particle.

High-energy particles in the RIAF radiate and lose energy by several processes. Electrons cool mainly via synchrotron emission and inverse Compton up-scattering of low-energy photons. The latter include those emitted by the thermal particles in the RIAF and the SSD as well as the synchrotron photons emitted by themselves (Synchrotron Self-Compton, SSC). The synchrotron cooling time for a charged particle of mass $m$ and Lorentz factor $\gamma$ moving in a medium with magnetic field $B$ is
\begin{equation}
    t_{\rm syn}(\gamma) = \frac{3 m c}{4\sigma_{\rm T} U_B} \left( \frac{m}{m_{\rm e}} \right)^2 \gamma^{-1},
\end{equation}
where $U_B=B^2/8\pi$ is the magnetic energy density.
Let $n_{\rm ph}(\epsilon)$ be the isotropically-averaged spectral density of low-energy photons (see Sect. \ref{sec:seds}), then the inverse Compton cooling timescale is \citep{moderski2005}
\begin{equation}
    t_{\rm IC}(\gamma)= \frac{3 m c}{4\sigma_{\rm T} U_{\rm ph}} \left( \frac{m}{m_{\rm e}} \right)^2 \gamma^{-1} 
    \left[ \frac{1}{U_{\rm ph}} \int d\epsilon ~\epsilon ~n_{\rm ph}(\epsilon)~ f_{\rm KN}(\tilde{b}) \right]^{-1},
\end{equation}
where $U_{\rm ph}$ is the photon energy density, and $f_{\rm KN}(\tilde{b})=9g(\tilde{b})/\tilde{b}^3$, where
\begin{align}
	g(\tilde{b}) = \left( \frac{1}{2} \tilde{b} + 6 + \frac{6}{\tilde{b}} \right) \ln (1+\tilde{b}) - \left( \frac{11}{12} \tilde{b}^3 + 6\tilde{b}^2+9\tilde{b}+4 \right) \\
	\times \frac{1}{\left( 1+\tilde{b}^2 \right)^2} -2 + 2 {\rm Li}_2 (-\tilde{b}).
\end{align}
Here, $\tilde{b}=4\gamma \epsilon/(m c^2)$.

In addition to the mechanisms mentioned above, relativistic protons lose energy through inelastic ${\rm pp}$ and ${\rm p}\gamma$ interactions. The cooling timescale due to ${\rm pp}$ collisions is
\begin{equation}
    t_{\rm pp}(\gamma)=\frac{1}{n_{\rm p} \sigma_{\rm pp} c \kappa_{\rm pp}},
\end{equation}
where $\sigma_{\rm pp}$ is the total cross-section and $\kappa_{\rm pp} \sim 0.17$ is the inelasticity of the process. An accurate parametrization for the total cross-section is given by \citep{kelner2006}
\begin{equation}
    \sigma_{\rm pp} \simeq \left( 34.3 + 1.88 L + 0.25 L^2 \right) \left[ 1 - \left(\frac{E_{\pi,{\rm thr}}}{E_{\rm p}}
    \right)^4 \right]^2~{\rm mb},
\end{equation}
for $E_{\rm p}>E_{\pi,{\rm thr}}$, where $E_{\rm p}=\gamma_{\rm p} m_{\rm p} c^2$ is the proton energy, $E_{\pi,{\rm thr}}=1.22~{\rm GeV}$ is the threshold energy for pion production, and $L=\log \left(E_{\rm p}/1~{\rm TeV}\right)$.
Photohadronic inelastic collisions cool protons via two channels: photomeson production (${\rm p}\gamma$) and photopair production---the so-called Bethe-Heitler (BH) channel. The cooling timescale for the first process is given by
\begin{equation}
    t_{{\rm p}\gamma}^{-1}=\frac{c}{2\gamma_{\rm p}} \int_{\bar{\varepsilon}_{{\rm thr},{\rm p}\gamma}}^\infty d\bar{\varepsilon} \sigma_{{\rm p}\gamma}\left(\bar{\varepsilon}\right)\kappa_{{\rm p}\gamma}(\bar{\varepsilon})\bar{\varepsilon} \int_{\bar{\varepsilon}/(2\gamma_{\rm p})}^\infty d\epsilon \frac{U_{\rm ph}(\epsilon)}{\epsilon^4},
    \label{eq:pgamma}
\end{equation}
where $\bar{\varepsilon}_{{\rm thr},{\rm p}\gamma}=145~{\rm MeV}$ is the threshold energy, $\sigma_{{\rm p}\gamma}$ is the cross-section, and $\kappa_{{\rm p}\gamma}$ is the inelasticity of the process. Useful parametrizations for these functions are given in \citet{atoyan2003}. For the Bethe-Heitler cooling channel, an equivalent expression is obtained by replacing the cross-section, inelasticity, and threshold energy in Eq. \ref{eq:pgamma} for their correspondent values $\sigma_{\rm BH}$, $\kappa_{\rm BH}$ (see, e.g., \citealt{begelman1990}), and $\bar{\varepsilon}_{\rm thr,BH}=2m_{\rm e}c^2$.

Both species of particles can escape from the system by two processes: energy-dependent diffusion in the vertical direction or standard RIAF winds.
The escape timescale via winds is independent of the particle energy and can be parameterized as
\begin{equation}
    t_{\rm wind}^{-1} = \frac{d \dot{ M}}{dr} \Bigg |_{\rm wind} \frac{v_r}{\dot{M}(r)} \approx s~\frac{|v_r|}{r}.
\end{equation}
The diffusion timescale is strongly dependent on the model adopted for the turbulence spectrum: $P(k) \propto k^{-q}$, where $k$ is the wavenumber. We assume a value of $q=5/3$, namely a Kolmogorov spectrum. Thus, the spatial diffusion coefficient for isotropically turbulent magnetic fields is
\begin{equation}
    D_r \approx \frac{c}{9\zeta}r_{\rm L} \left( k_{\rm min}r_{\rm L} \right)^{1-q},
\end{equation}
where $k_{\rm min}\sim H^{-1}$ is the minimum wave number of the turbulence spectrum, and the diffusive escape timescale is \citep{stawarz2008}
\begin{equation}
 t_{\rm diff} \approx \frac{H^2}{D_r} \simeq \frac {9H} {c} \zeta \left(\frac{r_L}{H}\right)^{q-2}\gamma^{q-2},
\end{equation}
where $\zeta= 8\pi\int P(k)dk/B_0^2$ is the strength ratio of turbulence fields against the background ordered component.
Finally, the total escape timescale is
\begin{equation}
t_{\rm esc} = \left( t_{\rm winds}^{-1} + t_{\rm diff}^{-1} \right)^{-1}.
\end{equation}

\begin{figure}
    \centering
    \includegraphics[width=\linewidth]{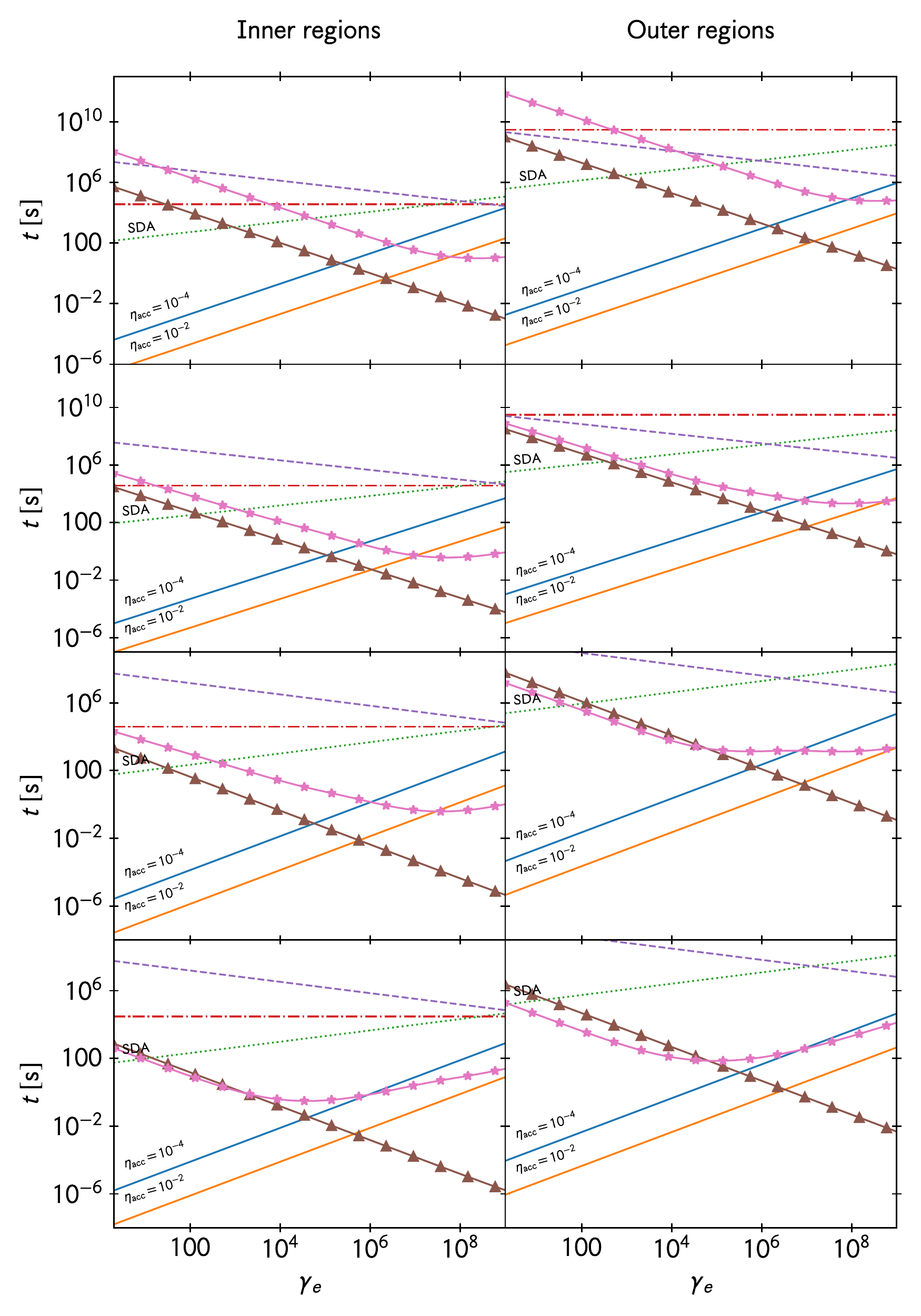}
    \caption{Timescales for nonthermal electrons in the various scenarios. Different rows show different scenarios (accretion rate increases downward), and each column shows different regions in the RIAF; left column: inner regions ($\sim 5 R_{\rm S}$), right column: outer regions ($\sim 200R_{\rm S}$). Plain solid lines show acceleration timescales, where a proper flag indicates the value of the acceleration efficiency $\eta_{\rm acc}$; dotted lines shows acceleration timescales for SDA (see Sect. \ref{subsec:acceleration}). Dashed lines show the diffusion timescale, dash-dotted lines show the advection timescale. Solid lines with markers show cooling timescales for the relevant processes; with triangles: synchrotron, with stars: inverse Compton.}
    \label{fig:timescales_e}
\end{figure}

\begin{figure}
    \centering
    \includegraphics[width=\linewidth]{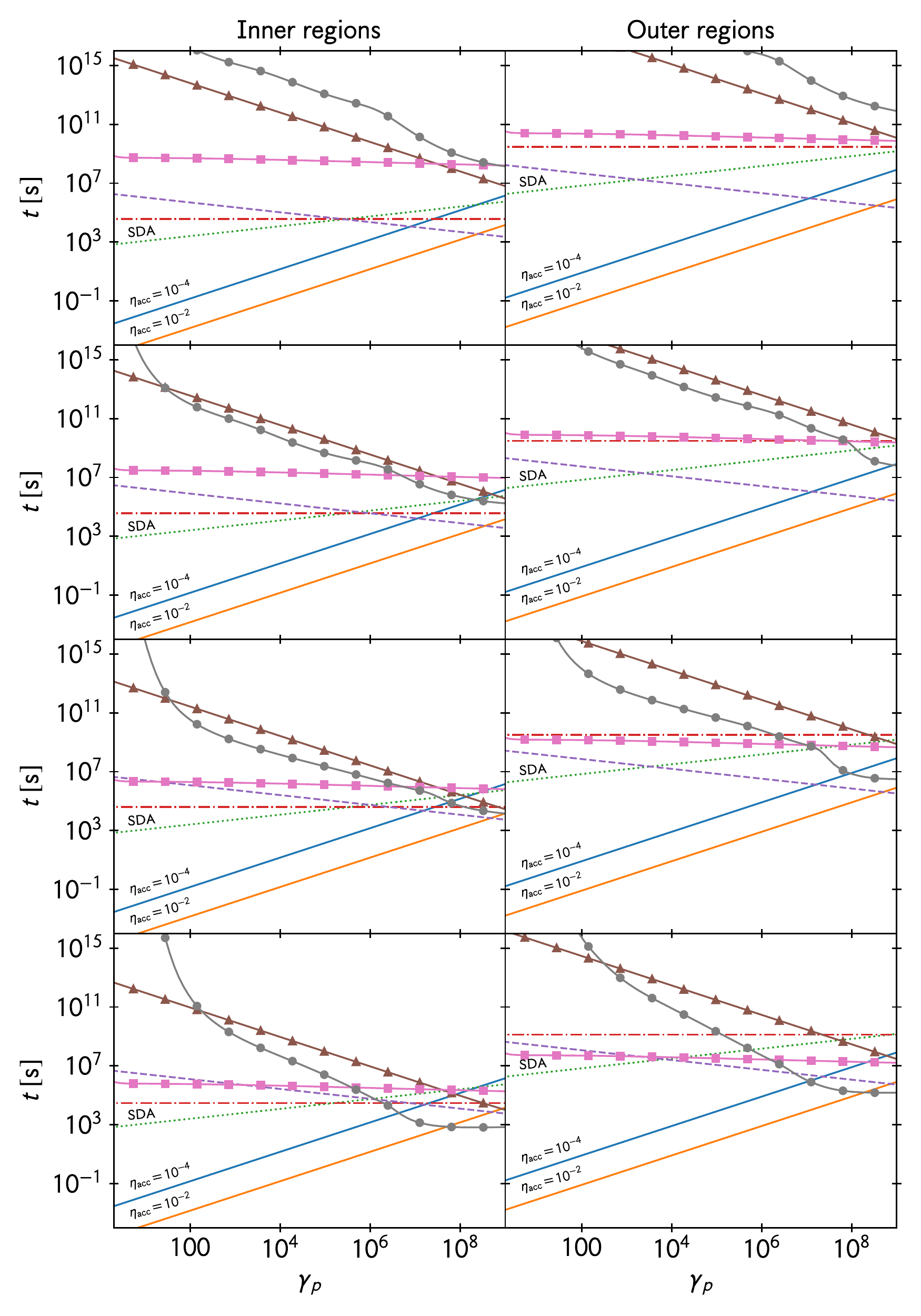}
    \caption{Same timescales as Fig. \ref{fig:timescales_e} but for protons. The cooling timescales shown are for synchrotron (triangles), ${\rm pp}$ (squares), and photohadronic losses (circles, includes ${\rm p}\gamma$ and BH).}
    \label{fig:timescales_p}
\end{figure}

Figure \ref{fig:timescales_e} shows acceleration, cooling, and escape timescales as a function of energy for the nonthermal electrons in the various scenarios considered. Each row shows a different set of parameters (accretion rate increasing downward), and the columns correspond to different regions in the RIAF; inner regions: $\sim 5R_{\rm S}$ and outer regions: $\sim 200R_{\rm S}$. The acceleration timescale is shown for two values of the acceleration efficiency: $\eta_{\rm acc}=10^{-2},10^{-4}$, and for the case of stochastic diffusive acceleration (SDA, see Sect. \ref{subsec:acceleration_2}). Synchrotron radiation and inverse Compton scattering are the dominant cooling processes for electrons and determine the maximum energy they can achieve. Inverse Compton becomes dominant at higher accretion rates, mainly because of the addition of copious amounts of seed photons emitted by the SSD penetrating the inner regions.

Figure \ref{fig:timescales_p} shows the same timescales as Fig. \ref{fig:timescales_e}, but for relativistic protons. At low accretion rates, the losses are completely dominated by escape processes, mainly by diffusion. At higher accretion rates, the magnetic field intensifies and advection competes with diffusion. Also, photohadronic losses start to be relevant and dominate at the highest energies as the RIAF luminosity and the photon density increase.

\subsection{Secondary particles}
\label{subsec:secondaries}

The hadronic processes described above lead to the production of secondary mesons and leptons. These particles will in turn emit radiation and suffer further interactions and decays \citep{reynoso2009}.
Inelastic ${\rm pp}$ and ${\rm p}\gamma$ collisions not only create neutral pions but also charged pions. The main channels for these processes are
\begin{flalign}
    {\rm p}+{\rm p} &\rightarrow {\rm p}+{\rm p} + \zeta_1 \pi^0 + \zeta_2(\pi^+ + \pi^-), \\
    {\rm p}+{\rm p} &\rightarrow {\rm p}+{\rm n} + \pi^+ + \zeta_1\pi^0 + \zeta_2(\pi^+ + \pi^-),
    \label{eq:pp_2}
\end{flalign}
and
\begin{flalign}
    {\rm p}+\gamma &\rightarrow {\rm p} + \zeta_1\pi^0 + \zeta_2(\pi^+ + \pi^-), \\
    {\rm p}+\gamma &\rightarrow {\rm n} + \pi^+ + \zeta_1\pi^0 + \zeta_2(\pi^+ + \pi^-),
    \label{eq:pgamma_2}
\end{flalign}
where $\zeta_1$ and $\zeta_2$ are the multiplicities. Charged pions have a mean lifetime of $\tau_{\pi^\pm}\simeq 2.6\times 10^{-8}~{\rm s}$ in their proper frame, and primarily decay into a muon and a muonic neutrino. The muon in turn decays with a mean lifetime of $\tau_{\mu^\pm}\simeq 2.2\times 10^{-6}~{\rm s}$ into an electron (or positron), an electronic neutrino, and a muon neutrino:
\begin{equation} \label{eq:pion_decay}
    \pi^\pm \rightarrow \mu^\pm + \nu_\mu (\bar{\nu}_\mu),
\end{equation}
\begin{equation} \label{eq:muon_decay}
    \mu^\pm \rightarrow {\rm e}^\pm + \bar{\nu}_\mu (\nu_\mu) + \nu_{\rm e}(\bar{\nu}_{\rm e}).
\end{equation}
Collisions between protons and photons also inject secondary pairs directly via the Bethe-Heitler (BH) process:
\begin{equation}
    {\rm p}+\gamma \rightarrow {\rm p} + {\rm e}^+ + {\rm e}^-.
\end{equation}
The last channel for the injection of secondary pairs is photon-photon annihilation. We include the presence of these particles and their interaction in our study. Useful approximations for the injection functions of pions and muons can be found in  \citet{atoyan2003,kelner2006,lipari2007}, whereas photopair production is studied in \citet{aharonian1983}.

\subsection{Coupled transport equations for the secondaries}
\label{subsec:secondaries_transport}

Secondary pairs are created with very high energies and they cool locally. Similarly, charged pions and muons decay or cool before being significantly transported in the radial direction. For simplicity, we track the evolution of secondary particles via the one-zone approximation at each radius in the RIAF. The evolution of pairs and photons is coupled through photo-annihilation absorption and must be solved iteratively. The coupled kinetic equations at the position $r$ for secondary pairs, charged pions, muons, and photons, respectively, are the following:

\begin{multline}
\label{eq:pairs}
		N_{{\rm e}^\pm}(\gamma_{{\rm e}^\pm}) \left[ t_{\rm cool}^{-1}+t_{\rm{diff}}^{-1} \right] = Q_{{\rm p}\gamma \rightarrow {\rm e}^\pm}(\gamma_{{\rm e}^\pm}) + 
		Q_{\mu^\pm \rightarrow {\rm e}^\pm}(\gamma_{{\rm e}^\pm}) \\ +  Q_{\gamma \gamma \rightarrow {\rm e}^\pm}(\gamma_{{\rm e}^\pm}).
\end{multline}
\begin{equation}\label{eq:pion}
		N_{\pi^\pm}(\gamma_\pi) \left[ t_{\rm cool}^{-1}+t_{\rm{diff}}^{-1} + t_{\rm dec}^{-1} \right] = Q_{{\rm p}\gamma \rightarrow \pi^\pm}(\gamma_\pi) + 
		Q_{{\rm pp} \rightarrow \pi^\pm}(\gamma_\pi).
\end{equation}
\begin{equation}\label{eq:muon}
		N_{\mu^\pm}(\gamma_\mu) \left[ t_{\rm cool}^{-1}+t_{\rm{diff}}^{-1} + t_{\rm dec}^{-1} \right] = Q_{\pi^\pm \rightarrow \mu^\pm}(\gamma_\mu).
	\end{equation} 
\begin{equation}\label{eq:photon}
	\begin{aligned}
		N_\gamma (\epsilon_\gamma)\left[ t_{\rm esc}^{-1} + t_{\gamma \gamma \rightarrow e^\pm}^{-1} \right] = Q_{\gamma}(\epsilon_\gamma),
		\end{aligned}
	\end{equation}

\noindent where $N_{\rm q}(\gamma_{\rm q})$ denotes the steady state particle energy distribution for the ${\rm q}$-species (in units of erg$^{-1}$cm$^{-3}$), and $t_{\rm dec} = \gamma_{\rm q} \tau_{\rm q}$ is its mean lifetime in the laboratory frame.
In Eq. \ref{eq:photon}, the term $Q_{\gamma}(E_{\gamma})$ includes the injection of photons by all radiative processes from every particle species.

\section{Spectral energy distributions}
\label{sec:seds}

Once the steady-state particle distributions are obtained, we calculate the radiative spectrum emitted by the primary and secondary particles. For thermal synchrotron, we use the parametrization of the emissivity given in \citet{mahadevan1996}. For electron-ion Bremsstrahlung we use the expression given in \citet{stepney1983}, with a small correction by \citet{narayan1995}, whereas we follow \citet{svensson1982} for electron-electron Bremsstrahlung; see also \citet{yarza2020} for a discussion on the different approximations used in the literature. To estimate the emission from the nonthermal populations, we use expressions for the emissivities per particle and integrate over the energy distribution. For the synchrotron emissivity we use the exact formula given in \citet{blumenthal1970}; we also take into account synchrotron-self absorption. For inverse Compton scattering, we assume an isotropic background photon field and use the formula given by \citet{moderski2005} (Eq. 16); for the neutral pion decay emissivity via hadronic interactions, we use the formalism of the delta-approximation as outlined in \citet{kelner2006} for ${\rm pp}$ collisions and the method described in \citet{kelner2008} for ${\rm p}\gamma$ collisions.

We follow \citet{manmoto1997}, and solve the radiative transfer in the vertical direction under the two-stream approximation \citep{rybicki1979}. The flux arising from each surface of the disk, without accounting for Comptonization, is
\begin{equation} \label{eq:flux}
    F_\nu = \frac{2\pi}{\sqrt{3}} S_\nu \left[ 1 - \exp \left(-2 \sqrt{3} \tau_\nu^* \right) \right],
\end{equation}
where $S_\nu = j_\nu / \kappa_\nu$ is the source function, $\tau_\nu^* = (\pi^{1/2} / 2) \kappa_\nu H$ is the half optical depth in the vertical direction, and $\kappa_\nu$ and $j_\nu$ are the absorption and emission coefficients. For the thermal spectrum $j_\nu = j_\nu^{\rm sync} + j_\nu^{\rm bremss}$, $\kappa_\nu = \kappa_\nu^{\rm th} = j_\nu / B_\nu$, and thus $S_\nu = B_\nu$.
The $N$ cylindrical shells are centered at $r_j$ and have boundaries at $l_{j-\frac{1}{2}}$ and $l_{j+\frac{1}{2}}$. Hence, the luminosity arising from the $j$-shell is
\begin{equation}
\label{eq:lum_flux}
    L_{\nu,j} = 2 \times \pi (l_{j+\frac{1}{2}}^2-l_{j-\frac{1}{2}}^2) \times F_{\nu,j},
\end{equation}
where the factor $2$ comes from the two faces of the disk.
In order to calculate the global coupling between different cells through Comptonization, we follow a similar approach to \citet{narayan1997}, namely we calculate scattering probability matrices coupling the different shells in the RIAF between themselves and with the cold disk, and we find iteratively the Comptonized luminosity as
\begin{equation}
    L_{\rm C,out}^k(\nu) = \int d\nu' \left( \frac{\nu}{\nu'} \right) L_{\rm C,in}^k(\nu') P(\nu,\nu',T_{\rm e}),
\end{equation}
where $P(\nu;\nu',T_{\rm e})$ is the probability for a photon of frequency $\nu'$ to be scattered to a frequency $\nu$ by and electron in a relativistic Maxwellian distribution of temperature $T_{\rm e}$ \citep{coppi1990}, and $L_{\rm C,in}^k$ is the luminosity emitted by all the shells in the RIAF and by the cold disk that reach the $k$-shell and gets scattered\footnote{We emphasize that to account for multiple Comptonization, this luminosity is updated at each iteration and in general includes previous orders of the Comptonization.}. We iterate until convergence (see \citealt{narayan1997} for more details).

For the nonthermal emission we restrict to local interactions and use Eqs. \ref{eq:flux} and \ref{eq:lum_flux} with the appropriate absorption and emission coefficients, namely $\kappa_\nu = \kappa_\nu^{\rm th} + \kappa_\nu^{\rm SSA}$ at low frequencies, and $\kappa_\nu^{\gamma \gamma} = \int d\epsilon ' n_{\rm ph}(\epsilon ') \sigma_{\gamma \gamma}(\epsilon_\gamma, \epsilon ')$ in the high-energy band accounting for internal photon-photon annihilation, where $\sigma_{\gamma \gamma}$ is the cross-section for the process \citep{gould1967}.
We include in the photon density $n_{\rm ph}$ both thermal and synchrotron nonthermal contributions. For the thermal photons, the most copious ones, we take into account nonlocal effects, and calculate the photon density in the shell centered at $r_j$ as
\begin{equation}
    n_{\rm ph}(\epsilon ) = \frac{L^{\rm NL}_j(\epsilon)}{\epsilon}~ \times \frac{t_{\rm cell}}{V_j},
\label{eq:n_ph}
\end{equation}
where $L_j^{\rm NL}(\epsilon)$ is the spectral luminosity emitted by all the shells in the RIAF that reach the shell $j$ (itself included). It is calculated via probability matrices that couple different shells in a similar way to how is done for Comptonization, and it includes redshift effects. The timescale $t_{\rm cell}$ is the average time a photon lives in shell $j$, and we estimate it as $\approx (H/c)~ \times ~ (1+\tau_{\rm es})$ for photons emitted in the shell $j$ and $\approx (l_{j+1}-l_{j})/c~ \times ~(1+\tau_{\rm es})$ for those coming from any other shell. Here, $\tau_{\rm es}=n_{\rm e} \sigma_{\rm T} H$ is the optical depth for Thomson scattering.
The nonthermal synchrotron photons are added to $n_{\rm ph}(\epsilon)$ at each shell and are calculated by Eq. \ref{eq:n_ph} though considering only local emission.
Finally, given the luminosity emitted by all the radiative processes at each shell, we calculate the total spectral luminosity measured by a distant observer as
\begin{equation}
    L_{\nu_{\rm o}} = \sum_j \frac{L_{\nu_{\rm e},j}}{\left[ 1 + z(r_j) \right]^3},
\end{equation}
where $\nu_{\rm e} = \left[ 1 + z(r) \right] \nu_{\rm o}$, and $z(r)= [(1-r^{-1})~(1-\beta_v^2)]^{-1/2}-1$ is the combined redshift accounting for both gravitational redshift and relativistic Doppler shift at the position $r$.

\section{Results}
\label{sec:results}

We have chosen the value of the power injected into protons such that for the models with $p=2$, $p_{\rm CR} \lesssim 0.2$ at all radii. In the inner regions, however, $p_{\rm CR}$ turns out to be $\ll 0.2$. This is explained because of two facts: Radial diffusion occurs mainly outward, and the prescription we choose for the injected power at different radii is fairly conservative. It is plausible that acceleration is much more efficient in the inner regions than in the outer ones.
The models with $p=1.2$ have even lower values of $p_{\rm CR}$ because more particles at higher energies imply more efficient cooling and diffusive escape.
The power injected in electrons is less constrained; we choose it in such a way that synchrotron emission does not heavily overtake the background thermal spectrum.

As discussed in Secs. \ref{sec:model_parameters} and \ref{subsec:acceleration}, we have chosen four accretion regimes, and two spectral slopes for particle injection. In turn, for the harder spectrum, we consider two values for particle acceleration efficiency: $\eta_{\rm acc}=10^{-4}$ as in the $p=2$ case, and a more efficient scenario with $\eta_{\rm acc}=10^{-2}$. Table \ref{table1} summarizes the main parameters of the different models. Here, $\varepsilon_{\rm NT}$ and $\varepsilon_{\rm e,p}$ are chosen as described above.

\begin{table}[ht]
    \caption[]{Main parameters adopted in the models.}
   	\label{table1}
   	\centering
\begin{tabular}{lccccccc}
\hline\hline %
Model & $\dot{m}_{\rm out}$ & $r_{\rm tr}$ & $b$ & $p$ & $\eta_{\rm acc}$ & $\varepsilon_{\rm NT}$ [$\%$] & $\varepsilon_{\rm e}$  \\ 
\hline
A1  & $10^{-4}$& $-$      & $-$   & $2$   & $10^{-4}$ & $0.5$ & $10^{-4}$ \\
A2 & $10^{-4}$ & $-$      & $-$   & $1.2$ & $10^{-4}$ & $0.5$ & $10^{-4}$ \\
A3 & $10^{-4}$ & $-$      & $-$   & $1.2$ & $10^{-2}$ & $0.5$ & $10^{-4}$ \\
B1 & $10^{-3}$ & $100$    & $2$   & $2$   & $10^{-4}$ & $1$   & $10^{-3}$ \\
B2 & $10^{-3}$ & $100$    & $2$   & $1.2$ & $10^{-4}$ & $1$   & $10^{-3}$ \\
B3 & $10^{-3}$ & $100$    & $2$   & $1.2$ & $10^{-2}$ & $1$   & $10^{-3}$ \\
C1 & $10^{-2}$ & $30$     & $1$   & $2$   & $10^{-4}$ & $1$   & $10^{-3}$ \\
C2 & $10^{-2}$ & $30$     & $1$   & $1.2$ & $10^{-4}$ & $1$   & $10^{-3}$ \\
C3 & $10^{-2}$ & $30$     & $1$   & $1.2$ & $10^{-2}$ & $1$   & $10^{-3}$ \\
D1 & $0.1$     & $3$      & $0.5$ & $2$   & $10^{-4}$ & $0.6$ & $1.6\times 10^{-2}$ \\
D2 & $0.1$     & $3$      & $0.5$ & $1.2$ & $10^{-4}$ & $0.6$ & $1.6\times 10^{-2}$ \\
D3 & $0.1$     & $3$      & $0.5$ & $1.2$ & $10^{-2}$ & $0.6$ & $1.6\times 10^{-2}$ \\
\hline  \\
\end{tabular}
\end{table}

\subsection{Steady-state particle distributions}

\begin{figure}
    \centering
    \includegraphics[width=0.9\linewidth]{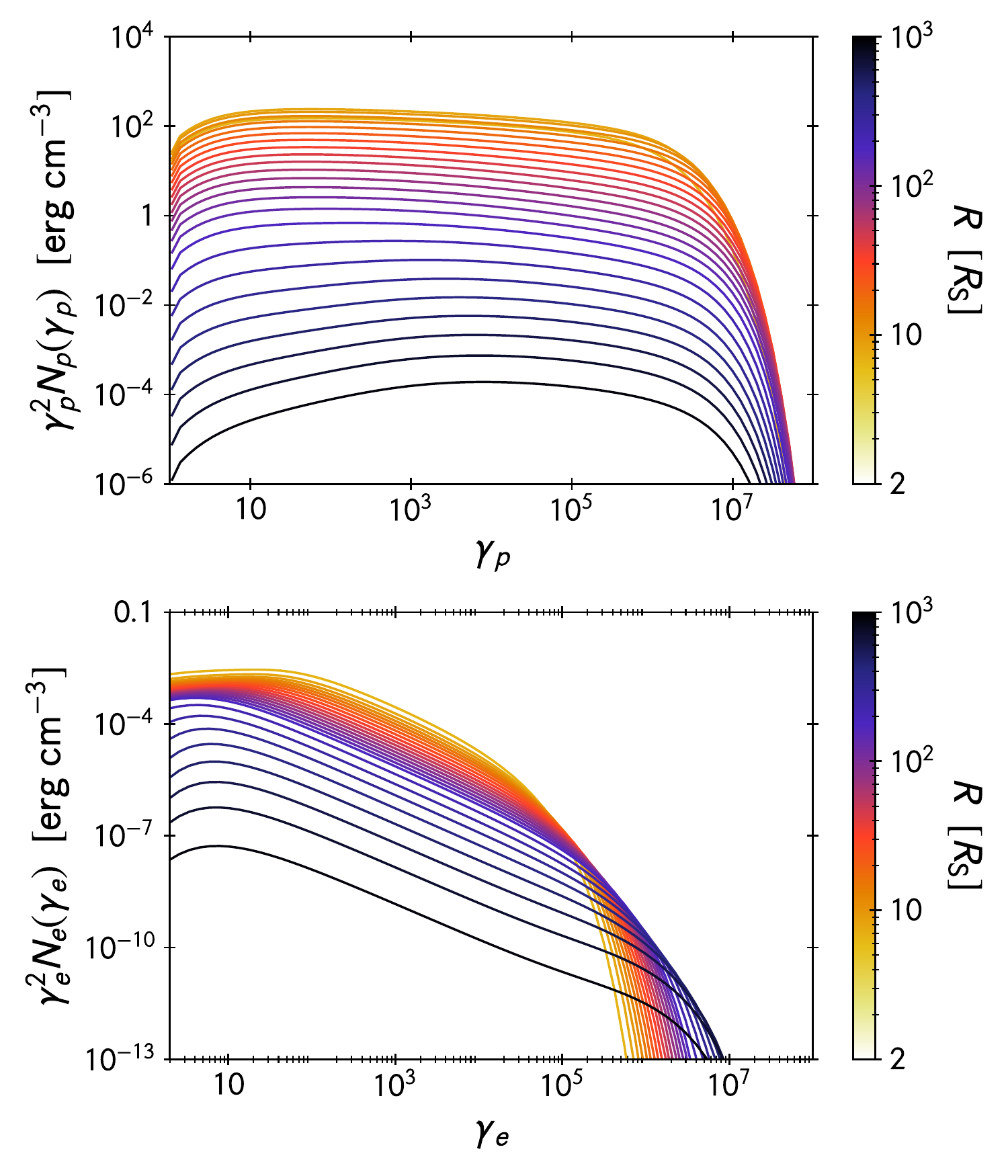}
    \caption{Steady particle energy distributions for Model B1 (see Table \ref{table1}). Different colored lines show different regions in the RIAF, which are indicated in the colorbar. Upper panel: protons. Lower panel: electrons.}
    \label{fig:distributions_3}
\end{figure}

\begin{figure}
    \centering
    \includegraphics[width=0.9\linewidth]{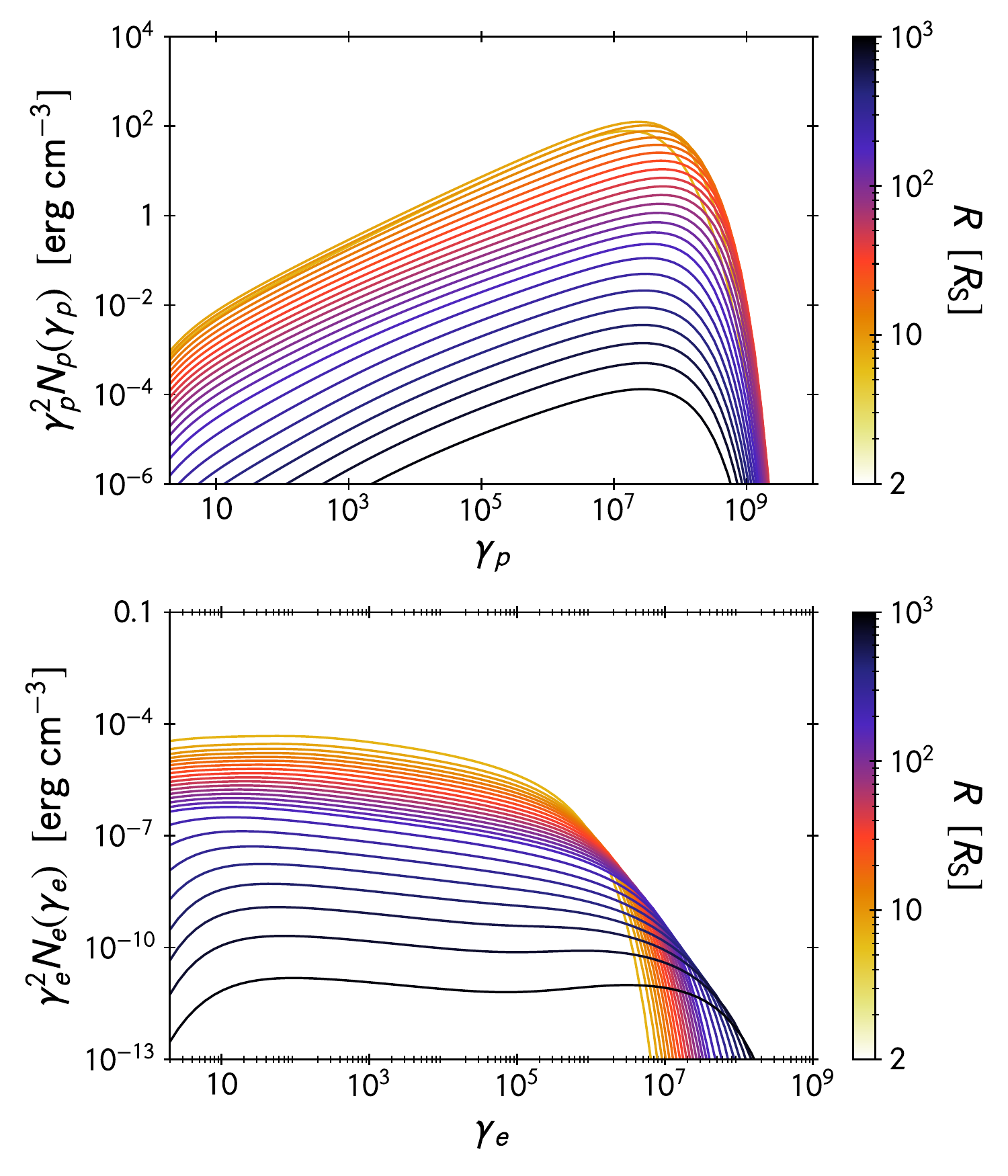}
    \caption{Steady particle energy distributions for Model B3 (see Table \ref{table1}). Different colored lines show different regions in the RIAF, which are indicated in the colorbar. Upper panel: protons. Lower panel: electrons.}
    \label{fig:distributions_3_hard}
\end{figure}

We show representative particle distributions of Models B1 and B3 in Figs. \ref{fig:distributions_3} and \ref{fig:distributions_3_hard}, respectively. Protons approximately maintain the injection spectral index, though little changes by the radial transport are seen in the outer layers. Electrons are cooled by synchrotron/inverse Compton mechanism. In the outer regions, a little hardening is seen at high energies, where inverse Compton scattering enters into the Klein-Nishina regime.

\subsection{Models A: Accretion rate $\dot{m}_{\rm out}=10^{-4}$}

\begin{figure*}
    \centering
    \includegraphics[width=0.95\linewidth]{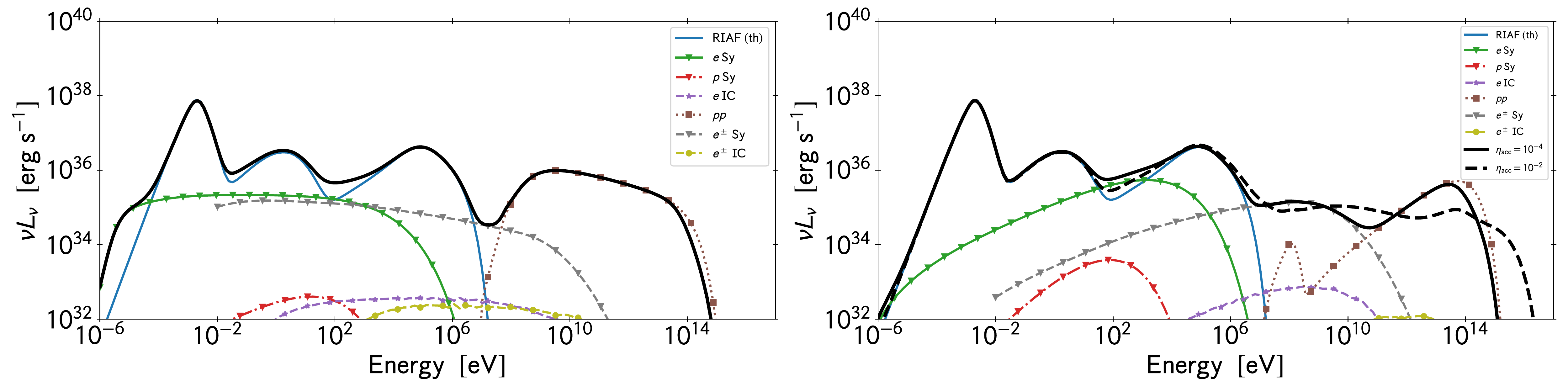}
    \caption{Accretion rate: $\dot{m}_{\rm out}=10^{-4}$. Left panel: Model A1. Right panel: Models A2 is shown in detail with the individual contributions and the absorbed total emission (solid dark line), and for Model A3 only the absorbed total emission (dashed dark line) is shown.}
    \label{fig:SED_4}
\end{figure*}

Models A correspond to the scenario with the lowest accretion rate ($\dot{m}_{\rm out}=10^{-4}$), where the flow is modeled as a pure RIAF at all radii; the calculated SEDs are shown in Fig. \ref{fig:SED_4}. The left panel shows Model A1, where nonthermal synchrotron emission is relatively strong, and a low electron power is enough to produce a bump in the radio band. If more power is injected into electrons, synchrotron emission will easily overcome the thermal luminosity. In the gamma-ray band, ${\rm pp}$ emission is also comparable to the broadband luminosity, but ${\rm p}\gamma$ emission is completely negligible due to the very low photon density. This low density implies that only the highest energy photons are absorbed. Electron-positron pairs are created mainly via muon decay and less by photo-annihilation, and their synchrotron emission is dominant in the megaelectronvolt band.

The right panel shows Models A2 and A3. A harder spectrum decreases the contribution of synchrotron emission to the radio band. Again, ${\rm pp}$ emission dominates in the very-high-energy band, but now the contribution of pairs is dominant in a broader range. The little bump in the ${\rm pp}$ spectrum at $\sim 100$ MeV corresponds to thermal protons. It is negligible because only very close to the event horizon protons reach temperatures high enough to create neutral pions, and the observed luminosity from those regions is heavily diminished due to gravitational and Doppler redshift. Nonetheless, this peak might be enhanced by beaming effects if the black hole rotates and we are seeing the disk at the proper inclination.

\subsection{Models B: Accretion rate $\dot{m}_{\rm out}=10^{-3}$, truncation radius $r_{\rm tr}=10^2$}

\begin{figure*}
    \centering
    \includegraphics[width=0.95\linewidth]{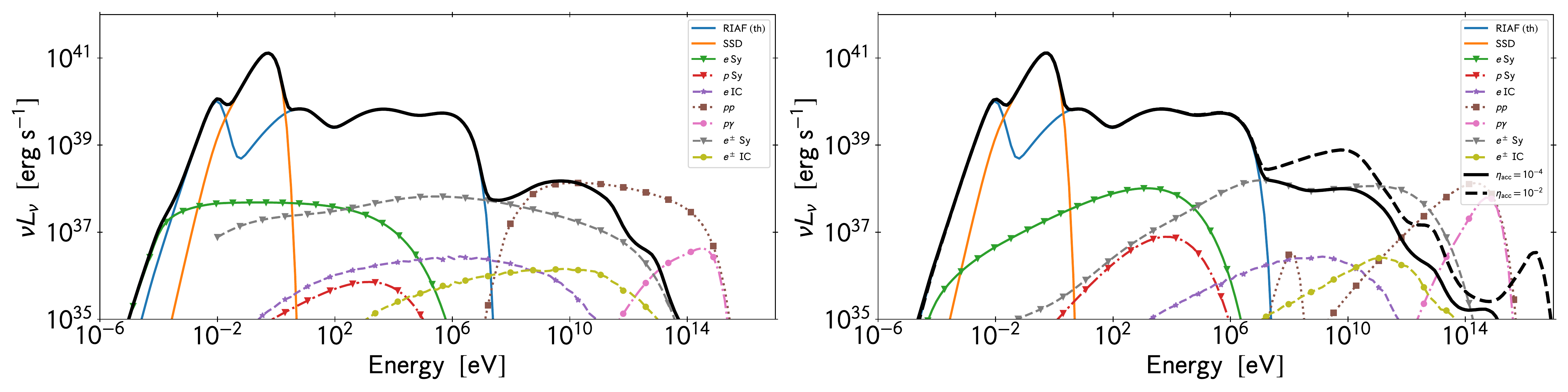}
    \caption{Accretion rate: $\dot{m}_{\rm out}=10^{-3}$. Left panel: Model B1. Right panel: Models B2 is shown in detail with the individual contributions and the absorbed total emission (solid dark line), and for Model B3 only the absorbed total emission (dashed dark line) is shown.}
    \label{fig:SED_3}
\end{figure*}

Models B correspond to the scenario with an accretion rate of $\dot{m}_{\rm out}=10^{-3}$. Here, we have considered an outer thin disk truncated at $r_{\rm tr}=100$; the calculated SEDs are shown in Fig. \ref{fig:SED_3}. The left panel shows in detail the contributions from the various processes in Model B1. Now, radio synchrotron emission is less notorious, and photo-annihilation starts to be important above $\sim 10$ GeV, due to the luminosity increase and the addition of the UV photons from the SSD. The contribution from the synchrotron emission by secondary pairs is similar to that in Models A.

The right panel shows the SEDs for Models B2 and B3. Now, ${\rm p}\gamma$ emission becomes comparable to ${\rm pp}$ emission in Model B2, and dominates in Model B3. This emission peaks at the PeV, and it is absorbed and reprocessed by the secondary pairs. The synchrotron emission of these secondaries dominates in the gamma-ray band and peaks at $\sim 10$ GeV.

\subsection{Models C: Accretion rate $\dot{m}_{\rm out}=10^{-2}$, truncation radius $r_{\rm tr}=30$}

\begin{figure*}
    \centering
    \includegraphics[width=0.95\linewidth]{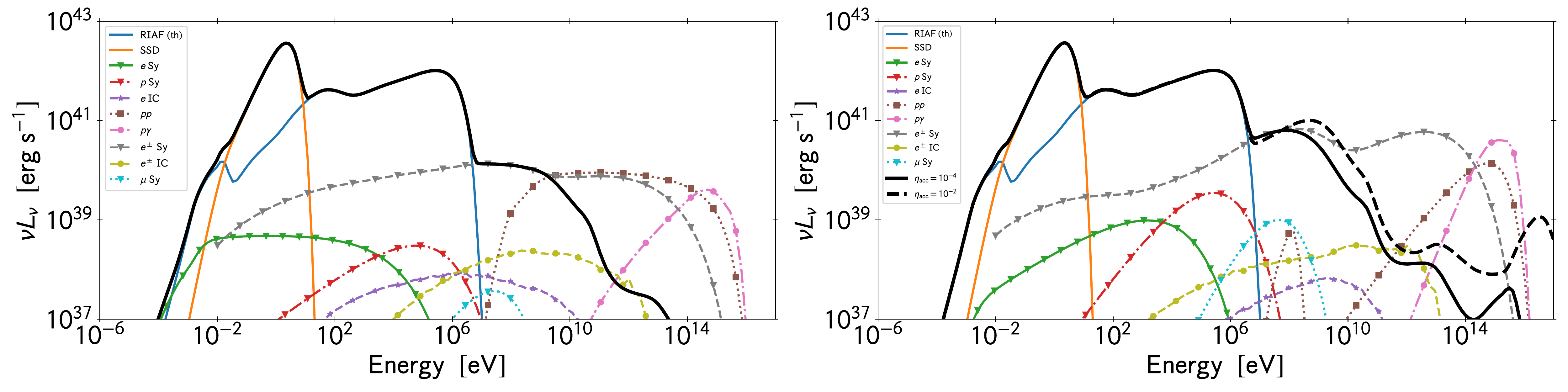}
    \caption{Accretion rate: $\dot{m}_{\rm out}=10^{-2}$. Left panel: Model C1. Right panel: Models C2 is shown in detail with the individual contributions and the absorbed total emission (solid dark line), and for Model C3 only the absorbed total emission (dashed dark line) is shown.}
    \label{fig:SED_2}
\end{figure*}

Models C correspond to the scenario with an accretion rate of $\dot{m}_{\rm out}=10^{-2}$. Here, we have considered an outer thin disk truncated at $r_{\rm tr}=30$; the calculated SEDs are shown in Fig. \ref{fig:SED_2}. The left panel shows in detail the contributions from the various processes in Model C1. Now, radio synchrotron emission is almost completely self-absorbed, and the gamma rays above the GeV are absorbed. The high-energy emission that escapes is synchrotron from secondary pairs.

The right panel shows the SEDs for Models C2 and C3. Now, ${\rm p}\gamma$ emission is higher than ${\rm pp}$ emission. The absorbed emission is reprocessed by the secondary pairs, whose synchrotron emission now peaks at $\sim 1$ GeV.

\subsection{Models D: Accretion rate $\dot{m}_{\rm out}=0.1$, truncation radius $r_{\rm tr}=3 \sim r_{\rm ISCO}$}

\begin{figure*}
    \centering
    \includegraphics[width=0.95\linewidth]{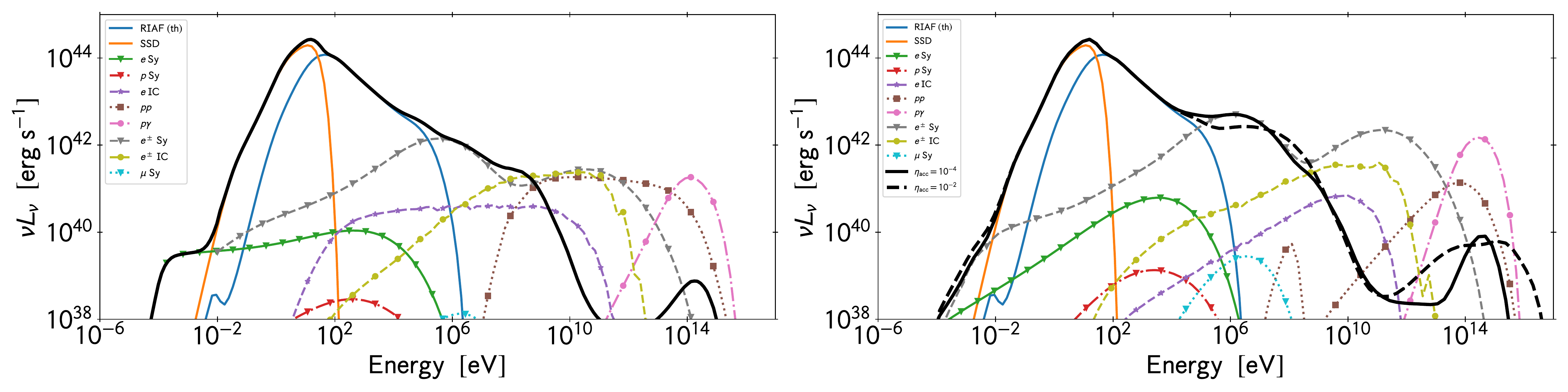}
    \caption{Accretion rate: $\dot{m}_{\rm out}=0.1$. Left panel: Model D1. Right panel: Models D2 is shown in detail with the individual contributions and the absorbed total emission (solid dark line), and for Model D3 only the absorbed total emission (dashed dark line) is shown.}
    \label{fig:SED_1}
\end{figure*}

Models D correspond to the scenario with an accretion rate of $\dot{m}_{\rm out}=0.1$. Here, we have considered an outer thin disk penetrating down the ISCO: $r_{\rm tr}=3$; the calculated SEDs are shown in Fig. \ref{fig:SED_1}. The left panel shows in detail the contributions from the various processes in Model D1. The physics of this flow is dominated by the emission from the SSD. The temperature of the corona is lower than in the other models because there are many more seed photons for inverse Compton cooling. Radio synchrotron emission is self-absorbed at lower energies and, again, produces a bump at $10$ GHz. The spectral index\footnote{$F_{\rm X} \propto E^{-\Gamma}$, where [$F_{\rm X}]=$ photons/keV/cm$^2$/s.} of the X-ray coronal emission depends on the transition parameter (see Sect. \ref{sec:IC4329A}) and, in this case, it is $\Gamma \approx -2.5$. The high-energy emission above the GeV is absorbed, and the reprocessed emission has a contribution from both synchrotron and inverse Compton pair emission. Inverse Compton from primary electrons now becomes more intense than synchrotron, but it is subdominant due to the low value of direct power going into electrons we have chosen. 

The right panel shows the SEDs for Models D2 and D3. Now, ${\rm p}\gamma$ emission is very strong, and the reprocessed radiation by secondary pairs produces a bump at energies above the megaelectronvolt range.

\section{An application: The corona in the Seyfert galaxy IC 4329A}
\label{sec:IC4329A}

\ic~is a bright X-ray Seyfert 1.2 galaxy\footnote{The class Seyfert 1.2 is used to describe objects with relatively weaker narrow H$\beta$ components, intermediate between Seyfert 1.0 and 1.5 \citep{veron2006}.} \citep{veron2006} at $z = 0.0161$ (\citealt{willmer1991}; equivalent to a luminosity distance of $69.61$~Mpc, assuming the cosmological parameters $H_0 = 70$ km s$^{-1}$Mpc$^{-1}$,  $\Omega_{\Lambda} = 0.7$, and $\Omega_{\rm m} = 0.3$). The mass of the central black hole is estimated in $\sim 1.2 \times 10^8 M_{\odot}$ \citep{markowitz2009,delacalle2010}. 
The host galaxy of the AGN is an edge-on spiral galaxy; the inclination of the disk of the host galaxy with respect to the axes of the AGN is thought to be the result of the interaction between \ic~and the companion galaxy IC~4329 at $\sim 3$ arcmin of separation \citep{Wolstencroft1995}.

The hard X-ray spectrum of the AGN is standard from a radio-quiet Seyfert, that is, a power-law resulting from the Compton up-scattering of the optically thick disk photons by the hot plasma of the RIAF or "corona". Although modest variability is observed, the power-law index in the Swift-BAT band ($14-195$ keV) is estimated as $\Gamma = 2.05^{+0.02}_{-0.03}$, and the total integrated flux is $F_{14-195{\rm keV}}=(263.25^{+2.9}_{-3.3})\times 10^{-12}~{\rm erg~s}^{-1}{\rm cm}^{-2}$, which at a distance of $69.61$ Mpc corresponds to a luminosity of $L_{14-195{\rm keV}}\simeq 10^{44.18}~{\rm erg~s}^{-1}$ \citep{oh2018}.

A moderated broadened Fe K$\alpha$ line has been reported by several authors, possibly indicating that the cold disk is truncated \citep{done2000}. Nevertheless, as discussed in \citealt{mantovani2014}, the nature of the emission line in this source is still under debate, and the high bolometric luminosity of the source seems to favor a radiatively efficient flow down to low radii. 

\begin{figure*}
    \centering
    \includegraphics[width=0.7\linewidth]{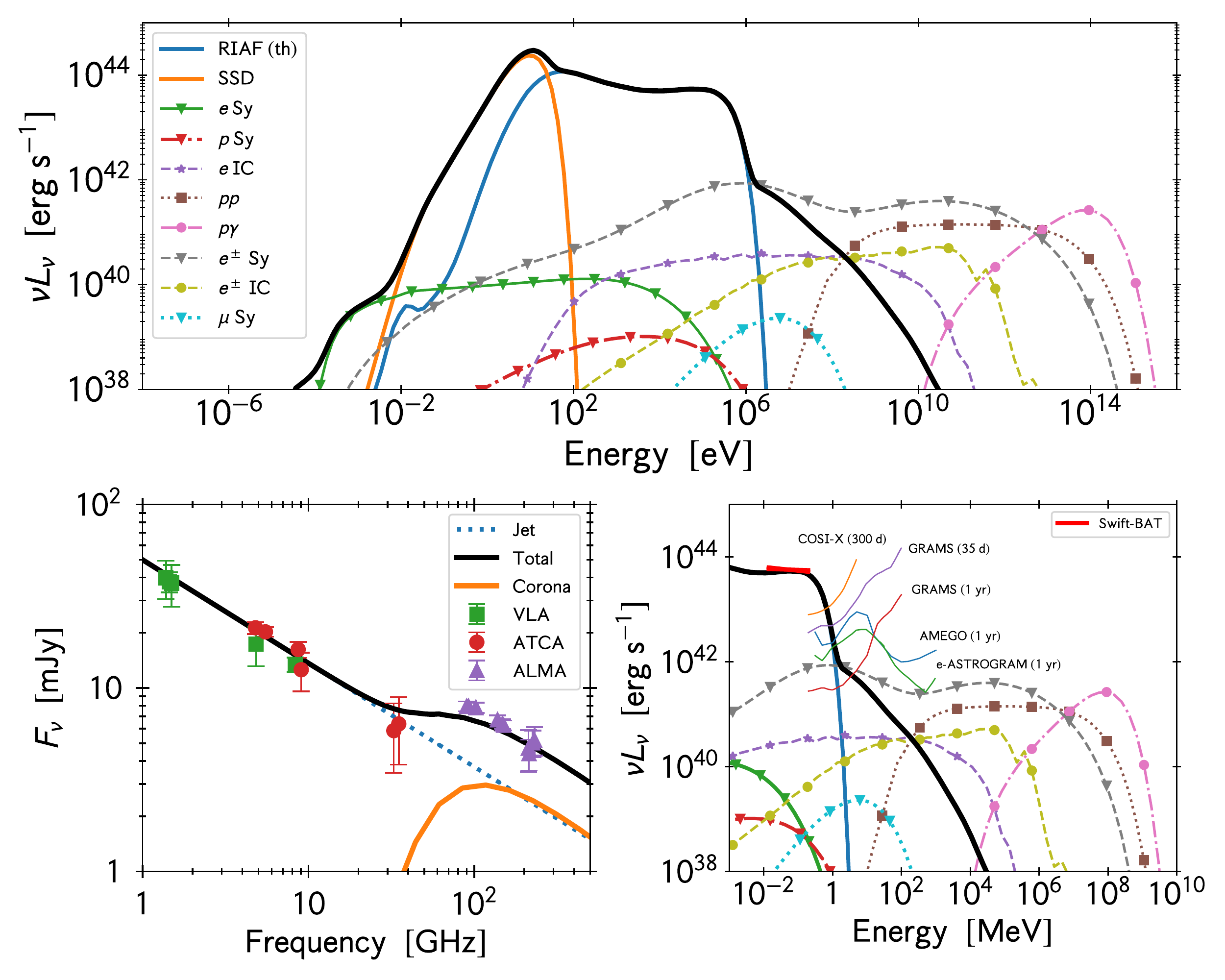}
    \caption{Seyfert galaxy IC $4329$A. The radio data are from \citet{inoue2018}, and the X-ray data are from \citet{oh2018}.  We also show the sensitivity of future MeV missions: COSI-X (300 days, \citealt{cosi2019}), e-ASTROGAM (1 year), GRAMS (35 days and 1 year, \citealt{grams2020}), and AMEGO (1 year, \citealt{amego2019}. For reference, we also included the sensitivity of three gamma-ray instruments: MAGIC (operating; above $100$~GeV), CTA (forthcoming; above $\sim 30$~GeV), and {\it Fermi} (operating; $\sim 0.1-100$~GeV).}
    \label{fig:IC4329A}
\end{figure*}

\citet{inoue2018} have shown that observations in the millimeter band of two Seyferts, one of them \ic, are well explained assuming nonthermal synchrotron emission in a hot corona. Additionally, inverse Compton emission by these electrons would contribute to the cosmic MeV background emission \citep{inoue2019}.
Figure \ref{fig:IC4329A} shows the SED predicted by our model. The parameters chosen are shown in Table \ref{tab:IC4329A}.
\begin{table*}
        \centering
        \caption{Parameters of our RIAF+thin disk model for the Seyfer galaxy \ic .}
        \begin{tabular}{l c}
                \hline\hline
                Parameter [units] & Value \\ \hline
                $m$ black hole mass [$\times 10^8$]&  $1.2$ \\
                $\dot{m}_{\rm out}$ outer accretion rate &  $0.12$ \\
                $r_{\rm out}$ outer radius &  $200$  \\
                $r_{\rm tr}$ truncation radius & $4$   \\
                $\alpha$ viscosity parameter &  $0.3$  \\
                $\beta$ plasma parameter & $10$ \\
                $\delta$ fraction of energy heating electrons &  $0.2$  \\
                $s$ wind parameter &  $0.1$  \\
                $b$ transition parameter & $0.1$  \\
                \hline
                $\varepsilon_{\rm NT}$ fraction of the accretion power going to nonthermal particles [$\%$] &  $5$ \\
                $\varepsilon_{\rm e}$ fraction of nonthermal power into electrons & $10^{-3}$ \\
                $\eta_{\rm acc}$ acceleration efficiency & $10^{-4}$ \\
                $p$ spectral index of injection & $2$ \\
                \hline\hline
        \end{tabular}
        \label{tab:IC4329A}
  \end{table*}
To model this source, \citet{inoue2019} assumed an homogeneous corona of radius $40$ $R_{\rm S}$, where the required magnetic field is of $\sim 10$ G. Our model predicts higher magnetic fields that increase toward the inner regions. We can explain the radio features as produced by synchrotron emission from nonthermal electrons in the outer layers of the corona. Moreover, since we considered the emission from the secondary particles, the contribution at high energies is different; in our model it is dominated by synchrotron radiation from secondary pairs, whereas in \citet{inoue2019} it is dominated by direct inverse Compton from primaries. \citet{inoue2018} already suggested that the corona in this system is likely an advection-heated hot accretion flow. This approach naturally explains the magnetic fields required for both efficient particle acceleration and synchrotron emission, and it is consistent with recent numerical results \citep{kimura2019b}. The inhomogeneous nature of the flow in our model also allows different regions of the flow with different magnetic fields and nonthermal power to produce the various features in the spectrum. Moreover, the secondary particle (hadronic in nature) origin for the presumed high-energy tail in the megaelectronvolt range is favored by recent particle-in-cell simulations \citep{zhdankin2019}.

\section{Discussion}
\label{sec:discussion}

\subsection{Particle acceleration mechanisms}
\label{subsec:acceleration_2}

Particle acceleration is not self-consistently treated in our model, but it is included via the injection functions of primary electrons and protons.
The most plausible acceleration mechanisms in hot collisionless RIAFs or coronae are magnetic reconnection and stochastic acceleration by magnetic turbulence.
A turbulent magnetized flow naturally gives rise to fast magnetic reconnection, since turbulence induces magnetic fluxes of opposite polarity to encounter each other at high velocities ($\sim V_{\rm A}$). Under these conditions, magnetic energy is transferred to the particles in the form of thermal, bulk and kinetic energy of individual particles. The latter involves particle acceleration (see \citealt{hoshino2013} for a review). Fast reconnection leads to efficient particle acceleration at a rate $t_{\rm acc}^{-1} \propto \gamma^{-a}$, with $0.2 < a < 0.6$, and power-law indices $N(\gamma) \propto \gamma^{-1,-2}$ \citep{delvalle2016,ball2018,werner2018}.

Magnetic reconnection also serves as a mechanism to push particles out from the thermal bath facilitating further stochastic acceleration via collisions between the particles and scattering centers produced by the turbulence.
SDA produces hard spectra that deviates from a simple power-law \citep{park1996,becker2006,kimura2015}, and the acceleration timescale differs from Eq. \ref{eq:acc_timescale}.
This timescale can be estimated as
\begin{equation}
    t_{\rm SDA} = \frac{\bar{p}^2}{D_{\bar{p}}},
\end{equation}
where $\bar{p}$ is the momentum of the particle and $D_{\bar{p}}$ is the diffusion coefficient in the momentum space; according to the quasi-linear theory \citep{dermer1996}, it is given by
\begin{equation}
    D_{\bar{p}} \simeq (mc)^2 (ck_{\rm min}) \left(\frac{v_{\rm A} }{c}\right)^2\zeta(r_{\rm L} k_{\rm min})^{j-2}\gamma^2,
\end{equation}
where $v_{\rm A} =B/\sqrt{4\pi \rho}$ is the Alfv\'en speed. Figures \ref{fig:timescales_e} and \ref{fig:timescales_p} include the SDA timescale, for which we have taken $\zeta=0.2$ and $j=5/3$ (Kolmogorov). This process is quite ineffective to accelerate electrons since they cool too fast. Protons cool much less efficient and are able to reach high energies, though lower than in our models.

Another acceleration process that has been considered in the literature is diffusive shock acceleration \citep{drury1983, inoue2019}. Nevertheless, this process requires that the plasma is compressible, and hence not highly magnetized ($\rho v^2 \gg B^2/8\pi$, see, e.g., \citealt{romero2018}). Since we are dealing with magnetized plasmas, the two quantities above are comparable and strong shocks are not expected to occur.

\subsection{Neutrino production}

Despite most of the high-energy electromagnetic emission is internally absorbed, neutrinos produced by photomeson interactions (see Eqs. \ref{eq:pion_decay} and \ref{eq:muon_decay}) escapes almost freely.
The study of neutrino production in accretion flows is of particular interest given that an excess of $2.9\sigma$ over the neutrino background was found in the ten-year survey data of IceCube, coincident with the direction of a nearby type-2 Seyfert galaxy \citep{aartsen2020}. \citet{inoue2020} studied the production of neutrinos in the corona of this source (see also \citealt{inoue2019}), and found that it can explain the excess, within a certain range of parameters in their model. 

\begin{figure}
    \centering
    \includegraphics[width=0.9\linewidth]{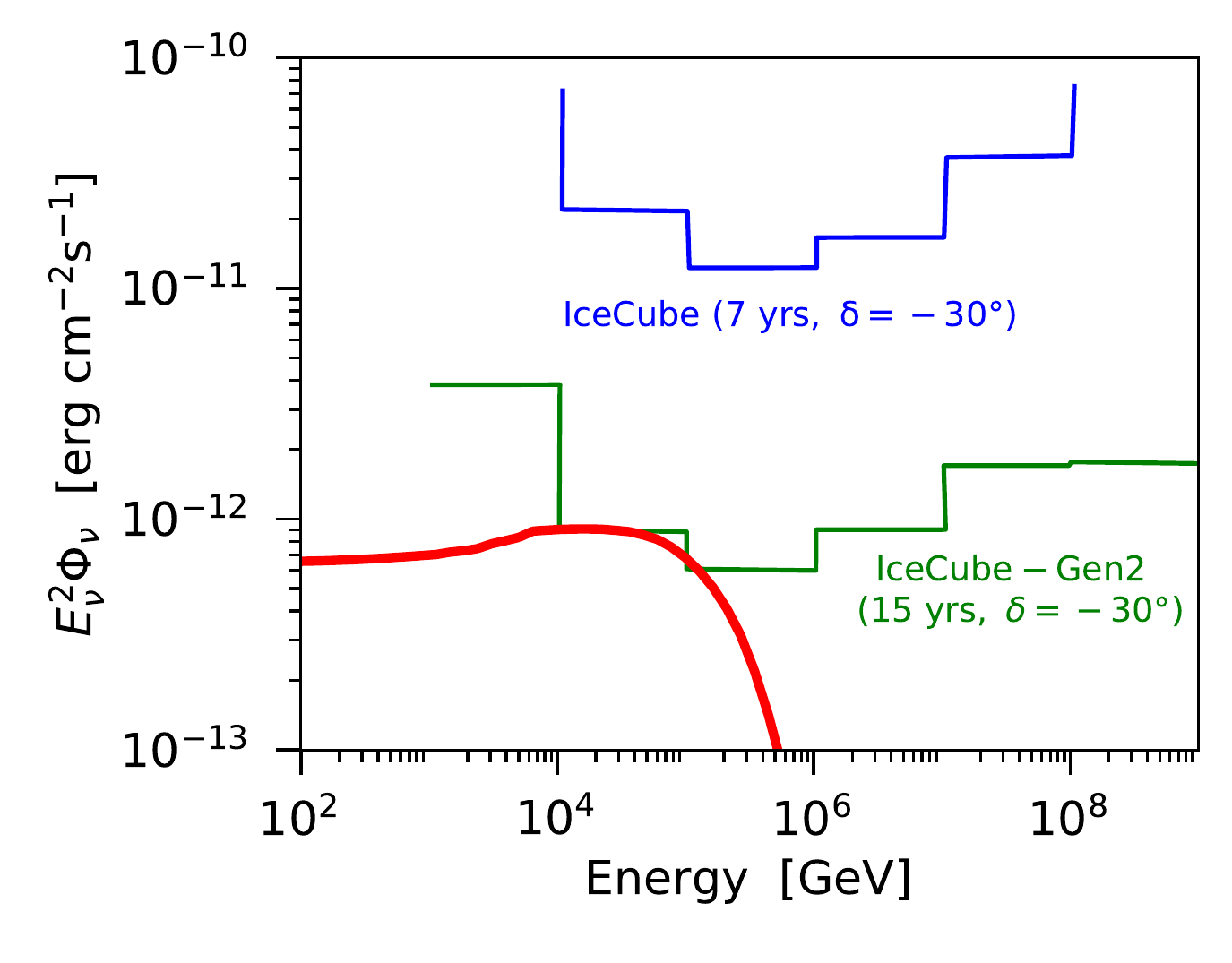}
    \caption{Total neutrino flux predicted for the Seyfert galaxy IC $4329$A. Sensitivity curves for IceCube and IceCube-Gen2 are shown \citep{vansanten2017,iceCube2019}}
    \label{fig:neutrino_IC4329A}
\end{figure}

A rough analysis of our results indicates that we should expect significant neutrino emission only for sources with moderate/high accretion rates, since, as discussed in Sect. \ref{sec:seds}, for low accretion rate  ${\rm p}\gamma$ is irrelevant and the luminosities achieved by ${\rm pp}$ interactions are low. We show in Figure \ref{fig:neutrino_IC4329A} the expected neutrino flux for \ic. Our hadronic-dominated model predicts a high neutrino emission that could be marginally detected in the future by IceCube-Gen2. The total contribution of the population of HAFs in the Universe to the neutrino background will be investigated in a future work.

\subsection{MeV background}

The soft-gamma ray extragalactic background ($\sim$1-10 MeV) is likely the result of the contribution from different sources, including SN Ia \citep{Ruiz_Lapuente_2016}, {\sl Fermi} blazars peaking in the MeV band (\citealt{giommi2015}, although most contribution from blazars is above 10 MeV), and also radio-quiet AGN if the accretion flow contains nonthermal particles \citep{stecker1999,inoue2007}. \citet{inoue2019} studied the contribution to the MeV background by hot accretion flows, and obtained that the measured fluxes can be explained as inverse Compton emission by nonthermal electrons in Seyfert coronae. The main difference with respect to our work is that they obtain primary electron fluxes to be dominant, hence cascades are not relevant. In our model, synchrotron emission from secondary pairs usually dominates over direct inverse Compton in the MeV band. This situation has also been found by other authors for the case of stochastic acceleration \citep{murase2020} and magnetic reconnection acceleration \citep{kheirandish2021}. In some extreme scenarios, proton and muon synchrotron emission can also contribute significantly in the megalectronvolt range (see, e.g., \citealt{romero2020}). The hadronic content is, then, an important component to be taken into account when studying the contribution of hot accretion flows to the MeV background.

\subsection{Neutron production and loading of jets}

Radiatively inefficient accretion flows are usually associated with the launching and collimation of relativistic jets in AGNs and microquasars. Jets launched by the Blandford-Znajek (BZ) process \citep{blandford1977} requires the accumulation of magnetic flux in the innermost regions close to the black hole ergosphere \citep{tchekhovskoy2011}, which is favored by the advective nature of HAFs. On the other hand, both the high $H/R$ ratio of the flow and the ubiquitous presence of wind in these systems help to collimate the jet at its first stage \citep{yuan2014}.
Since BZ jets are launched as purely Poynting fluxes, an important problem to deal with is how this electromagnetic outflow can be loaded with mass at the base of the jet. The presence of matter is inferred by very-high-resolution observations of the nearby AGN M87, which show that radiation associated with the jet is being produced at distances down to ~5 Schwarzschild radii from the central supermassive black hole \citep{hada2013, eht2019}.

The nonthermal processes in accretion flows onto black hole that we discussed may play a non-negligible role in loading jets with charged particles. 
The interaction of relativistic protons with matter (${\rm pp}$) and radiation (${\rm p}\gamma$) produce neutrons via the channels given by Eqs. \ref{eq:pp_2} and \ref{eq:pgamma_2}. These neutrons freely escape from the corona and decay into relativistic protons and electrons at long distances; a fraction of them will decay within the jet funnel. This mechanism was proposed as a means to load Poynting-dominated outflows with baryons \citep{toma2012,vila2014}. Moreover, neutrons may collide with photons creating pions through the process $n+\gamma \rightarrow {\rm p} + \pi^-$. Pion decay will quickly inject pairs in the funnel \citep{romero2020}.

Pairs can also be created via the annihilation of photons emitted in the corona; this process can be separated into two categories:
{\it a)} MeV-MeV collisions, which will take place even when nonthermal processes are not significant (thermal MeV photons), and {\it b)} collisions between a high-energy (nonthermal) gamma ray and a soft photon from the accretion disk. The latter mechanism is highly dependent on the details of the nonthermal mechanisms occurring in the corona (see \citealt{romero2020} for numerical estimates of the particle loading through these various mechanisms).

\subsection{Comparison with other works}

The semi-analytical approach we have taken has pros and cons when compared with detailed numerical simulations. As we mentioned, simulations are unavoidable to study nonlinear physics and multidimensional phenomena (like outflows) or complex time variability \citep{yuan2014}. Nevertheless, in many situations they require focusing on only a small part of the system (as it is the case in Particle-in-Cell simulations) or upon a few of all the relevant physical processes that take place: Nonthermal processes are usually neglected in magnetohydrodynamic simulations of accretion flows. Semi-analytical models, on the other hand, rely on their versatility. They allow to treat globally the flow, including several physical processes at different time or spatial scales, and the interpretation of the results is more direct.

Several previous semi-analytical approaches to investigate nonthermal processes in RIAFs were presented in the literature. \citet{kimura2015} developed a model considering the RIAF as a homogeneous spherical hot plasma where the physical fields are obtained either from the self-similar solution \citep{narayan1994} or from numerical simulations \citep{kimura2019}. They propose stochastic acceleration of protons as a means to produce nonthermal processes that could lead to multi-messenger outputs. Recently, \citet{inoue2019} developed a more observationally-motivated corona model based on the detection of nonthermal coronal activity in two nearby Seyfert galaxies; they also considered a homogeneous spherical flow. Our model presents improvements with respect to these works: We combine the detailed treatment of nonthermal processes with actual hydrodynamic solutions of hot accretion flows. The latter provides a radial dependence of the physical fields in the flow, which in turn permits to calculate a more accurate thermal background. In addition, we included self-consistently the presence of a cold thin disk coexisting with the hot flow. Over this background, we introduced the presence of a population of nonthermal particles and study both the spatial and energy transport, the production of secondary particles, and calculate all the relevant outputs.

\section{Conclusions}
\label{sec:conclusions}

We have developed a detailed model to study nonthermal processes occurring in hot accretion flows. The model is flexible enough to be applied to a broad range of accretion rates and luminosities. It consists of a hot accretion flow, modeled as an RIAF, plus a cold thin disk. For various sets of parameters, we investigated the most relevant nonthermal processes that occur, including the particle transport and the nonthermal radiative output. 

In models with accretion rate $\dot{m}> 10^{-3}$, the radiation above GeV is highly attenuated due to self-absorption, hence all emission in this domain is expected to originate in jets. At lower accretion rates (Models A) most of the radiation can escape. On the other hand, the contribution from both thermal and nonthermal RIAF emission to the MeV/sub-GeV band might be dominant in the local Universe.

We also applied our model to the source \ic, confirming results from previous works, in which millimetric observations of this source can be well reproduced by synchrotron emission in a hot corona. In our model, the corona is heated by the magnetorotational instability, and the millimetric excess comes from further distances to the hole than those considered by the previous authors \citep{inoue2018}.

 Our model presents some improvements to previous models of nonthermal processes in HAFs:
\begin{itemize}
    \item[$\bullet$] By solving the actual hydrodynamics equations for a hot accretion flow, we obtain a description of the flow that takes into account the radial dependence of the fluid properties, and where the values of these properties arise self-consistently from the solution itself.\\
    
    \item[$\bullet$] We include the presence of a thin disk that coexists and interacts with the hot accretion flow, as it is expected to occur in many AGNs.\\
    
    \item[$\bullet$] The thermal radiative emission is calculated with great detail, taking into account nonlocal effects, and it is consistent with the fluid properties. This, together with the two previous improvements, set up a more realistic background over which nonthermal processes are calculated.\\
    
    \item[$\bullet$] The nonthermal transport includes the spatial advection and diffusion of particles when these processes are important, namely for protons. Besides, all relevant secondary processes are calculated.\\
\end{itemize}

The transition from thermal to nonthermal emission from RIAFs would take place in the MeV band of the SED. Henceforth, it is fundamental to cover this energy range to better understand particle acceleration in these systems. Besides, studies on gamma-ray polarization can be useful to disentangle the origin of the radiation (coronae vs jets, or starbursts when present), as it was used in the case of BHBs \citep{laurent2011}. After the successful COMPTEL instrument, on board the Compton Gamma-Ray Observatory (1991-2000), we lack a detector in the MeV band. Future MeV gamma-ray missions, such as GRAMS \citep{grams2020}, and in particular, those with polarimetric facilities, as AMEGO ($0.2$ MeV$ - 10$ GeV \citealt{amego2019}), and COSI-X ($0.2-5$ MeV, \citealt{cosi2019}), might shed light on these subjects, and put to the test our model.

\begin{acknowledgements}
We thank the referee, D. Khangulyan, for useful comments that helped to improve the manuscript.
We also thank Kohta Murase and Ali Kheirandish for helpful observations about our work and Enrico Peretti for valuable discussions about the transport of relativistic particles.
This work was supported by the Argentine agency CONICET (PIP 2014-00338), the National Agency for Scientific and Technological Promotion (PICT 2017-0898 and PICT 2017-2865). G.E.R. acknowledges the support by the Spanish Ministerio de Ciencia e Innovaci\'{o}n (MICINN) under grant PID2019-105510GB-C31 and through the ``Center of
Excellence Mar\'ia de Maeztu 2020-2023'' award to the ICCUB (CEX2019-000918-M).
\end{acknowledgements}

%
\bibliographystyle{aa} 
\bibliography{biblio} 

\begin{thebibliography}{127}
\expandafter\ifx\csname natexlab\endcsname\relax\def\natexlab#1{#1}\fi

\bibitem[{{Aartsen} {et~al.}(2020){Aartsen}, {Ackermann}, {Adams}, {Aguilar},
  {Ahlers}, {Ahrens}, {Alispach}, {Andeen}, {Anderson}, {Ansseau}, {Anton},
  {Arg{\"u}elles}, {Auffenberg}, {Axani}, {Backes}, \&
  {Bagherpour}}]{aartsen2020}
{Aartsen}, M.~G., {Ackermann}, M., {Adams}, J., {et~al.} 2020, \prl, 124,
  051103

\bibitem[{{Aartsen} {et~al.}(2019){Aartsen}, {Ackermann}, {Adams}, {Aguilar},
  {Ahlers}, {Ahrens}, {Alispach}, {Andeen}, {Anderson}, {Ansseau}, \&
  et~al.}]{iceCube2019}
{Aartsen}, M.~G., {Ackermann}, M., {Adams}, J., {et~al.} 2019, arXiv e-prints,
  arXiv:1911.02561

\bibitem[{{Abdollahi} {et~al.}(2020){Abdollahi}, {Acero}, {Ackermann},
  {Ajello}, {Atwood}, {Axelsson}, {Baldini}, {Ballet}, {Barbiellini},
  {Bastieri}, {Becerra Gonzalez}, {Bellazzini}, {Berretta}, {Bissaldi}, {Bland
  ford}, {Bloom}, {Bonino}, {Bottacini}, {Brandt}, {Bregeon}, {Bruel},
  {Buehler}, {Burnett}, {Buson}, {Cameron}, {Caputo}, {Caraveo}, {Casandjian},
  {Castro}, {Cavazzuti}, {Charles}, {Chaty}, {Chen}, {Cheung}, {Chiaro},
  {Ciprini}, {Cohen-Tanugi}, {Cominsky}, {Coronado-Bl{\'a}zquez}, {Costantin},
  {Cuoco}, {Cutini}, {D'Ammando}, {DeKlotz}, {de la Torre Luque}, {de Palma},
  {Desai}, {Digel}, {Di Lalla}, {Di Mauro}, {Di Venere}, {Dom{\'\i}nguez},
  {Dumora}, {Fana Dirirsa}, {Fegan}, {Ferrara}, {Franckowiak}, {Fukazawa},
  {Funk}, {Fusco}, {Gargano}, {Gasparrini}, {Giglietto}, {Giommi}, {Giordano},
  {Giroletti}, {Glanzman}, {Green}, {Grenier}, {Griffin}, {Grondin}, {Grove},
  {Guiriec}, {Harding}, {Hayashi}, {Hays}, {Hewitt}, {Horan},
  {J{\'o}hannesson}, {Johnson}, {Kamae}, {Kerr}, {Kocevski}, {Kovac'evic'},
  {Kuss}, {Landriu}, {Larsson}, {Latronico}, {Lemoine-Goumard}, {Li},
  {Liodakis}, {Longo}, {Loparco}, {Lott}, {Lovellette}, {Lubrano}, {Madejski},
  {Maldera}, {Malyshev}, {Manfreda}, {Marchesini}, {Marcotulli},
  {Mart{\'\i}-Devesa}, {Martin}, {Massaro}, {Mazziotta}, {McEnery}, {Mereu},
  {Meyer}, {Michelson}, {Mirabal}, {Mizuno}, {Monzani}, {Morselli},
  {Moskalenko}, {Negro}, {Nuss}, {Ojha}, {Omodei}, {Orienti}, {Orlando},
  {Ormes}, {Palatiello}, {Paliya}, {Paneque}, {Pei}, {Pe{\~n}a-Herazo},
  {Perkins}, {Persic}, {Pesce-Rollins}, {Petrosian}, {Petrov}, {Piron}, {Poon},
  {Porter}, {Principe}, {Rain{\`o}}, {Rando}, {Razzano}, {Razzaque}, {Reimer},
  {Reimer}, {Remy}, {Reposeur}, {Romani}, {Saz Parkinson}, {Schinzel},
  {Serini}, {Sgr{\`o}}, {Siskind}, {Smith}, {Spandre}, {Spinelli}, {Strong},
  {Suson}, {Tajima}, {Takahashi}, {Tak}, {Thayer}, {Thompson}, {Tibaldo},
  {Torres}, {Torresi}, {Valverde}, {Van Klaveren}, {van Zyl}, {Wood},
  {Yassine}, \& {Zaharijas}}]{Abdollahi2020}
{Abdollahi}, S., {Acero}, F., {Ackermann}, M., {et~al.} 2020, \apjs, 247, 33

\bibitem[{{Abramowicz} {et~al.}(1988){Abramowicz}, {Czerny}, {Lasota}, \&
  {Szuszkiewicz}}]{abramowicz1988}
{Abramowicz}, M.~A., {Czerny}, B., {Lasota}, J.~P., \& {Szuszkiewicz}, E. 1988,
  \apj, 332, 646

\bibitem[{{Aharonian} {et~al.}(1983){Aharonian}, {Atoyan}, \&
  {Nagapetyan}}]{aharonian1983}
{Aharonian}, F.~A., {Atoyan}, A.~M., \& {Nagapetyan}, A.~M. 1983, Astrophysics,
  19, 187

\bibitem[{{Aharonian} {et~al.}(2002){Aharonian}, {Belyanin}, {Derishev},
  {Kocharovsky}, \& {Kocharovsky}}]{aharonian2002}
{Aharonian}, F.~A., {Belyanin}, A.~A., {Derishev}, E.~V., {Kocharovsky}, V.~V.,
  \& {Kocharovsky}, V.~V. 2002, \prd, 66, 023005

\bibitem[{{Aramaki} {et~al.}(2020){Aramaki}, {Adrian}, {Karagiorgi}, \&
  {Odaka}}]{grams2020}
{Aramaki}, T., {Adrian}, P. O.~H., {Karagiorgi}, G., \& {Odaka}, H. 2020,
  Astroparticle Physics, 114, 107

\bibitem[{{Atoyan} \& {Dermer}(2003)}]{atoyan2003}
{Atoyan}, A.~M. \& {Dermer}, C.~D. 2003, \apj, 586, 79

\bibitem[{{Ball} {et~al.}(2018){Ball}, {Sironi}, \& {{\"O}zel}}]{ball2018}
{Ball}, D., {Sironi}, L., \& {{\"O}zel}, F. 2018, \apj, 862, 80

\bibitem[{{Bandyopadhyay} {et~al.}(2019){Bandyopadhyay}, {Xie}, {Nagar},
  {Schleicher}, {Ramakrishnan}, {Ar{\'e}valo}, {L{\'o}pez}, \&
  {Diaz}}]{Bandyopadhyay2019}
{Bandyopadhyay}, B., {Xie}, F.-G., {Nagar}, N.~M., {et~al.} 2019, \mnras, 490,
  4606

\bibitem[{{Becker} {et~al.}(2006){Becker}, {Le}, \& {Dermer}}]{becker2006}
{Becker}, P.~A., {Le}, T., \& {Dermer}, C.~D. 2006, \apj, 647, 539

\bibitem[{{Begelman}(2014)}]{begelman2014}
{Begelman}, M.~C. 2014, arXiv e-prints, arXiv:1410.8132

\bibitem[{{Begelman} {et~al.}(1990){Begelman}, {Rudak}, \&
  {Sikora}}]{begelman1990}
{Begelman}, M.~C., {Rudak}, B., \& {Sikora}, M. 1990, \apj, 362, 38

\bibitem[{{Bisnovatyi-Kogan} \& {Blinnikov}(1977)}]{Bisnovatyi1977}
{Bisnovatyi-Kogan}, G.~S. \& {Blinnikov}, S.~I. 1977, \aap, 59, 111

\bibitem[{{Blandford} \& {Eichler}(1987)}]{blandford1987}
{Blandford}, R. \& {Eichler}, D. 1987, \physrep, 154, 1

\bibitem[{{Blandford} \& {Begelman}(1999)}]{blandford1999}
{Blandford}, R.~D. \& {Begelman}, M.~C. 1999, \mnras, 303, L1

\bibitem[{{Blandford} \& {Znajek}(1977)}]{blandford1977}
{Blandford}, R.~D. \& {Znajek}, R.~L. 1977, \mnras, 179, 433

\bibitem[{{Blumenthal} \& {Gould}(1970)}]{blumenthal1970}
{Blumenthal}, G.~R. \& {Gould}, R.~J. 1970, Reviews of Modern Physics, 42, 237

\bibitem[{{Chael} {et~al.}(2017){Chael}, {Narayan}, \& {Sadowski}}]{chael2017}
{Chael}, A.~A., {Narayan}, R., \& {Sadowski}, A. 2017, \mnras, 470, 2367

\bibitem[{{Chandrasekhar}(1939)}]{chandrasekhar1939}
{Chandrasekhar}, S. 1939, {An introduction to the study of stellar structure}

\bibitem[{{Chang} \& {Cooper}(1970)}]{chang-cooper1970}
{Chang}, J.~S. \& {Cooper}, G. 1970, Journal of Computational Physics, 6, 1

\bibitem[{{Chen} {et~al.}(1995){Chen}, {Abramowicz}, {Lasota}, {Narayan}, \&
  {Yi}}]{chen1995}
{Chen}, X., {Abramowicz}, M.~A., {Lasota}, J.-P., {Narayan}, R., \& {Yi}, I.
  1995, \apjl, 443, L61

\bibitem[{{Chiang} \& {Blaes}(2003)}]{chiang2003}
{Chiang}, J. \& {Blaes}, O. 2003, \apj, 586, 97

\bibitem[{{Coppi}(1992)}]{coppi1992}
{Coppi}, P.~S. 1992, \mnras, 258, 657

\bibitem[{{Coppi} \& {Blandford}(1990)}]{coppi1990}
{Coppi}, P.~S. \& {Blandford}, R.~D. 1990, \mnras, 245, 453

\bibitem[{{de Gouveia Dal Pino} {et~al.}(2010){de Gouveia Dal Pino},
  {Piovezan}, \& {Kadowaki}}]{degouveiadalpino2010}
{de Gouveia Dal Pino}, E.~M., {Piovezan}, P.~P., \& {Kadowaki}, L.~H.~S. 2010,
  \aap, 518, A5

\bibitem[{{de La Calle P{\'e}rez} {et~al.}(2010){de La Calle P{\'e}rez},
  {Longinotti}, {Guainazzi}, {Bianchi}, {Dov{\v{c}}iak}, {Cappi}, {Matt},
  {Miniutti}, {Petrucci}, {Piconcelli}, {Ponti}, {Porquet}, \&
  {Santos-Lle{\'o}}}]{delacalle2010}
{de La Calle P{\'e}rez}, I., {Longinotti}, A.~L., {Guainazzi}, M., {et~al.}
  2010, \aap, 524, A50

\bibitem[{{del Valle} {et~al.}(2016){del Valle}, {de Gouveia Dal Pino}, \&
  {Kowal}}]{delvalle2016}
{del Valle}, M.~V., {de Gouveia Dal Pino}, E.~M., \& {Kowal}, G. 2016, \mnras,
  463, 4331

\bibitem[{{Dermer} {et~al.}(1991){Dermer}, {Liang}, \& {Canfield}}]{dermer1991}
{Dermer}, C.~D., {Liang}, E.~P., \& {Canfield}, E. 1991, \apj, 369, 410

\bibitem[{{Dermer} {et~al.}(1996){Dermer}, {Miller}, \& {Li}}]{dermer1996}
{Dermer}, C.~D., {Miller}, J.~A., \& {Li}, H. 1996, \apj, 456, 106

\bibitem[{{Dexter} {et~al.}(2020){Dexter}, {Tchekhovskoy},
  {Jim{\'e}nez-Rosales}, {Ressler}, {Baub{\"o}ck}, {Dallilar}, {de Zeeuw},
  {Eisenhauer}, {von Fellenberg}, {Gao}, {Genzel}, {Gillessen}, {Habibi},
  {Ott}, {Stadler}, {Straub}, \& {Widmann}}]{dexter2020}
{Dexter}, J., {Tchekhovskoy}, A., {Jim{\'e}nez-Rosales}, A., {et~al.} 2020,
  \mnras [\eprint[arXiv]{2006.03657}]

\bibitem[{{Done} {et~al.}(2000){Done}, {Madejski}, \& {{\.Z}ycki}}]{done2000}
{Done}, C., {Madejski}, G.~M., \& {{\.Z}ycki}, P.~T. 2000, \apj, 536, 213

\bibitem[{{Dove} {et~al.}(1997){Dove}, {Wilms}, \& {Begelman}}]{dove1997}
{Dove}, J.~B., {Wilms}, J., \& {Begelman}, M.~C. 1997, \apj, 487, 747

\bibitem[{{Drury}(1983)}]{drury1983}
{Drury}, L.~O. 1983, Reports on Progress in Physics, 46, 973

\bibitem[{{Esin} {et~al.}(1997){Esin}, {McClintock}, \& {Narayan}}]{esin1997}
{Esin}, A.~A., {McClintock}, J.~E., \& {Narayan}, R. 1997, \apj, 489, 865

\bibitem[{{Esin} {et~al.}(1998){Esin}, {Narayan}, {Cui}, {Grove}, \&
  {Zhang}}]{esin1998}
{Esin}, A.~A., {Narayan}, R., {Cui}, W., {Grove}, J.~E., \& {Zhang}, S.-N.
  1998, \apj, 505, 854

\bibitem[{{Esin} {et~al.}(1996){Esin}, {Narayan}, {Ostriker}, \&
  {Yi}}]{esin1996}
{Esin}, A.~A., {Narayan}, R., {Ostriker}, E., \& {Yi}, I. 1996, \apj, 465, 312

\bibitem[{{Event Horizon Telescope Collaboration} {et~al.}(2019){Event Horizon
  Telescope Collaboration}, {Akiyama}, {Alberdi}, {Alef}, {Asada}, {Azulay},
  {Baczko}, {Ball}, {Balokovi{\'c}}, {Barrett}, \& et~al.}]{eht2019}
{Event Horizon Telescope Collaboration}, {Akiyama}, K., {Alberdi}, A., {et~al.}
  2019, \apjl, 875, L1

\bibitem[{{Frank} {et~al.}(2002){Frank}, {King}, \& {Raine}}]{frank2002}
{Frank}, J., {King}, A., \& {Raine}, D.~J. 2002, {Accretion Power in
  Astrophysics: Third Edition}

\bibitem[{{Giommi} \& {Padovani}(2015)}]{giommi2015}
{Giommi}, P. \& {Padovani}, P. 2015, \mnras, 450, 2404

\bibitem[{{Gould} \& {Schr{\'e}der}(1967)}]{gould1967}
{Gould}, R.~J. \& {Schr{\'e}der}, G.~P. 1967, Physical Review, 155, 1404

\bibitem[{{Guti{\'e}rrez} {et~al.}(2020){Guti{\'e}rrez}, {Nemmen}, \&
  {Cafardo}}]{gutierrez2020a}
{Guti{\'e}rrez}, E.~M., {Nemmen}, R., \& {Cafardo}, F. 2020, \apjl, 891, L36

\bibitem[{{Hada} {et~al.}(2013){Hada}, {Kino}, {Doi}, {Nagai}, {Honma},
  {Hagiwara}, {Giroletti}, {Giovannini}, \& {Kawaguchi}}]{hada2013}
{Hada}, K., {Kino}, M., {Doi}, A., {et~al.} 2013, \apj, 775, 70

\bibitem[{{Hilburn} {et~al.}(2010){Hilburn}, {Liang}, {Liu}, \&
  {Li}}]{hilburn2010}
{Hilburn}, G., {Liang}, E., {Liu}, S., \& {Li}, H. 2010, \mnras, 401, 1620

\bibitem[{{Hoshino}(2013)}]{hoshino2013}
{Hoshino}, M. 2013, \apj, 773, 118

\bibitem[{{Hoshino} \& {Lyubarsky}(2012)}]{hoshino2012}
{Hoshino}, M. \& {Lyubarsky}, Y. 2012, \ssr, 173, 521

\bibitem[{Inoue \& Doi(2018)}]{inoue2018}
Inoue, Y. \& Doi, A. 2018, The Astrophysical Journal, 869, 114

\bibitem[{{Inoue} {et~al.}(2020){Inoue}, {Khangulyan}, \& {Doi}}]{inoue2020}
{Inoue}, Y., {Khangulyan}, D., \& {Doi}, A. 2020, \apjl, 891, L33

\bibitem[{{Inoue} {et~al.}(2019){Inoue}, {Khangulyan}, {Inoue}, \&
  {Doi}}]{inoue2019}
{Inoue}, Y., {Khangulyan}, D., {Inoue}, S., \& {Doi}, A. 2019, \apj, 880, 40

\bibitem[{Inoue {et~al.}(2007)Inoue, Totani, \& Ueda}]{inoue2007}
Inoue, Y., Totani, T., \& Ueda, Y. 2007, \apj, 672, L5

\bibitem[{{Kelner} \& {Aharonian}(2008)}]{kelner2008}
{Kelner}, S.~R. \& {Aharonian}, F.~A. 2008, \prd, 78, 034013

\bibitem[{{Kelner} {et~al.}(2006){Kelner}, {Aharonian}, \&
  {Bugayov}}]{kelner2006}
{Kelner}, S.~R., {Aharonian}, F.~A., \& {Bugayov}, V.~V. 2006, \prd, 74, 034018

\bibitem[{{Kheirandish} {et~al.}(2021){Kheirandish}, {Murase}, \&
  {Kimura}}]{kheirandish2021}
{Kheirandish}, A., {Murase}, K., \& {Kimura}, S.~S. 2021, arXiv e-prints,
  arXiv:2102.04475

\bibitem[{{Kimura} {et~al.}(2019{\natexlab{a}}){Kimura}, {Murase}, \&
  {M{\'e}sz{\'a}ros}}]{kimura2019}
{Kimura}, S.~S., {Murase}, K., \& {M{\'e}sz{\'a}ros}, P. 2019{\natexlab{a}},
  \prd, 100, 083014

\bibitem[{{Kimura} {et~al.}(2015){Kimura}, {Murase}, \& {Toma}}]{kimura2015}
{Kimura}, S.~S., {Murase}, K., \& {Toma}, K. 2015, \apj, 806, 159

\bibitem[{{Kimura} {et~al.}(2019{\natexlab{b}}){Kimura}, {Tomida}, \&
  {Murase}}]{kimura2019b}
{Kimura}, S.~S., {Tomida}, K., \& {Murase}, K. 2019{\natexlab{b}}, \mnras, 485,
  163

\bibitem[{{Laurent} {et~al.}(2011){Laurent}, {Rodriguez}, {Wilms}, {Cadolle
  Bel}, {Pottschmidt}, \& {Grinberg}}]{laurent2011}
{Laurent}, P., {Rodriguez}, J., {Wilms}, J., {et~al.} 2011, Science, 332, 438

\bibitem[{{Li} \& {Miller}(1997)}]{li1997}
{Li}, H. \& {Miller}, J.~A. 1997, \apjl, 478, L67

\bibitem[{{Lipari} {et~al.}(2007){Lipari}, {Lusignoli}, \&
  {Meloni}}]{lipari2007}
{Lipari}, P., {Lusignoli}, M., \& {Meloni}, D. 2007, \prd, 75, 123005

\bibitem[{{Liu} \& {Wu}(2013)}]{liu2013}
{Liu}, H. \& {Wu}, Q. 2013, \apj, 764, 17

\bibitem[{{Lynn} {et~al.}(2014){Lynn}, {Quataert}, {Chandran}, \&
  {Parrish}}]{lynn2014}
{Lynn}, J.~W., {Quataert}, E., {Chandran}, B. D.~G., \& {Parrish}, I.~J. 2014,
  \apj, 791, 71

\bibitem[{{Mahadevan} {et~al.}(1997){Mahadevan}, {Narayan}, \&
  {Krolik}}]{mahadevan1997a}
{Mahadevan}, R., {Narayan}, R., \& {Krolik}, J. 1997, \apj, 486, 268

\bibitem[{{Mahadevan} {et~al.}(1996){Mahadevan}, {Narayan}, \&
  {Yi}}]{mahadevan1996}
{Mahadevan}, R., {Narayan}, R., \& {Yi}, I. 1996, \apj, 465, 327

\bibitem[{{Mahadevan} \& {Quataert}(1997)}]{mahadevan1997b}
{Mahadevan}, R. \& {Quataert}, E. 1997, \apj, 490, 605

\bibitem[{{Manmoto} {et~al.}(1997){Manmoto}, {Mineshige}, \&
  {Kusunose}}]{manmoto1997}
{Manmoto}, T., {Mineshige}, S., \& {Kusunose}, M. 1997, \apj, 489, 791

\bibitem[{{Mantovani} {et~al.}(2014){Mantovani}, {Nandra}, \&
  {Ponti}}]{mantovani2014}
{Mantovani}, G., {Nandra}, K., \& {Ponti}, G. 2014, \mnras, 442, L95

\bibitem[{{Maraschi} \& {Tavecchio}(2003)}]{maraschi2003}
{Maraschi}, L. \& {Tavecchio}, F. 2003, \apj, 593, 667

\bibitem[{{Markowitz}(2009)}]{markowitz2009}
{Markowitz}, A. 2009, \apj, 698, 1740

\bibitem[{{McEnery} {et~al.}(2019){McEnery}, {van der Horst}, {Dominguez},
  {Moiseev}, {Marcowith}, {Harding}, {Lien}, {Giuliani}, {Inglis}, {Ansoldi},
  {Stamerra}, {Manousakis}, {Strong}, \& {Bambi}}]{amego2019}
{McEnery}, J., {van der Horst}, A., {Dominguez}, A., {et~al.} 2019, in Bulletin
  of the American Astronomical Society, Vol.~51, 245

\bibitem[{{Moderski} {et~al.}(2005){Moderski}, {Sikora}, {Coppi}, \&
  {Aharonian}}]{moderski2005}
{Moderski}, R., {Sikora}, M., {Coppi}, P.~S., \& {Aharonian}, F. 2005, \mnras,
  363, 954

\bibitem[{{Murase} {et~al.}(2020){Murase}, {Kimura}, \&
  {M{\'e}sz{\'a}ros}}]{murase2020}
{Murase}, K., {Kimura}, S.~S., \& {M{\'e}sz{\'a}ros}, P. 2020, \prl, 125,
  011101

\bibitem[{{Narayan}(1996)}]{narayan1996}
{Narayan}, R. 1996, \apj, 462, 136

\bibitem[{{Narayan} {et~al.}(1997){Narayan}, {Barret}, \&
  {McClintock}}]{narayan1997}
{Narayan}, R., {Barret}, D., \& {McClintock}, J.~E. 1997, \apj, 482, 448

\bibitem[{{Narayan} {et~al.}(1998){Narayan}, {Mahadevan}, {Grindlay}, {Popham},
  \& {Gammie}}]{narayan1998}
{Narayan}, R., {Mahadevan}, R., {Grindlay}, J.~E., {Popham}, R.~G., \&
  {Gammie}, C. 1998, \apj, 492, 554

\bibitem[{{Narayan} \& {Yi}(1994)}]{narayan1994}
{Narayan}, R. \& {Yi}, I. 1994, \apjl, 428, L13

\bibitem[{{Narayan} \& {Yi}(1995)}]{narayan1995}
{Narayan}, R. \& {Yi}, I. 1995, \apj, 444, 231

\bibitem[{{Nemmen} {et~al.}(2014){Nemmen}, {Storchi-Bergmann}, \&
  {Eracleous}}]{nemmen2014}
{Nemmen}, R.~S., {Storchi-Bergmann}, T., \& {Eracleous}, M. 2014, \mnras, 438,
  2804

\bibitem[{{Novikov} \& {Thorne}(1973)}]{novikov1973}
{Novikov}, I.~D. \& {Thorne}, K.~S. 1973, in Black Holes (Les Astres Occlus),
  343--450

\bibitem[{{Oh} {et~al.}(2018){Oh}, {Koss}, {Markwardt}, {Schawinski},
  {Baumgartner}, {Barthelmy}, {Cenko}, {Gehrels}, {Mushotzky}, {Petulante},
  {Ricci}, {Lien}, \& {Trakhtenbrot}}]{oh2018}
{Oh}, K., {Koss}, M., {Markwardt}, C.~B., {et~al.} 2018, \apjs, 235, 4

\bibitem[{{Oka} \& {Manmoto}(2003)}]{oka2003}
{Oka}, K. \& {Manmoto}, T. 2003, \mnras, 340, 543

\bibitem[{{{\"O}zel} {et~al.}(2000){{\"O}zel}, {Psaltis}, \&
  {Narayan}}]{ozel2000}
{{\"O}zel}, F., {Psaltis}, D., \& {Narayan}, R. 2000, \apj, 541, 234

\bibitem[{{Paczy{\'n}sky} \& {Wiita}(1980)}]{paczynsky1980}
{Paczy{\'n}sky}, B. \& {Wiita}, P.~J. 1980, \aap, 500, 203

\bibitem[{{Park} \& {Petrosian}(1996)}]{park1996}
{Park}, B.~T. \& {Petrosian}, V. 1996, \apjs, 103, 255

\bibitem[{{Peng} {et~al.}(2019){Peng}, {Zhang}, {Wang}, {Wang}, \&
  {Zhi}}]{peng2019}
{Peng}, F.-K., {Zhang}, H.-M., {Wang}, X.-Y., {Wang}, J.-F., \& {Zhi}, Q.-J.
  2019, \apj, 884, 91

\bibitem[{{Poutanen}(1998)}]{poutanen1998}
{Poutanen}, J. 1998, in Theory of Black Hole Accretion Disks, ed. M.~A.
  {Abramowicz}, G.~{Bj{\"o}rnsson}, \& J.~E. {Pringle}, 100--122

\bibitem[{{Poutanen} {et~al.}(1997){Poutanen}, {Krolik}, \&
  {Ryde}}]{poutanen1997}
{Poutanen}, J., {Krolik}, J.~H., \& {Ryde}, F. 1997, \mnras, 292, L21

\bibitem[{{Quataert} \& {Gruzinov}(1999)}]{quataert1999}
{Quataert}, E. \& {Gruzinov}, A. 1999, \apj, 520, 248

\bibitem[{{Reynoso} \& {Romero}(2009)}]{reynoso2009}
{Reynoso}, M.~M. \& {Romero}, G.~E. 2009, \aap, 493, 1

\bibitem[{{Rieger}(2017)}]{Rieger2017}
{Rieger}, F.~M. 2017, in American Institute of Physics Conference Series, Vol.
  1792, 6th International Symposium on High Energy Gamma-Ray Astronomy, 020008

\bibitem[{{Rodr{\'\i}guez-Ram{\'\i}rez}
  {et~al.}(2019){Rodr{\'\i}guez-Ram{\'\i}rez}, {de Gouveia Dal Pino}, \& {Alves
  Batista}}]{rodriguezramirez2019}
{Rodr{\'\i}guez-Ram{\'\i}rez}, J.~C., {de Gouveia Dal Pino}, E.~M., \& {Alves
  Batista}, R. 2019, \apj, 879, 6

\bibitem[{{Romero} \& {Guti{\'e}rrez}(2020)}]{romero2020}
{Romero}, G.~E. \& {Guti{\'e}rrez}, E.~M. 2020, Universe, 6, 99

\bibitem[{{Romero} {et~al.}(2018){Romero}, {M{\"u}ller}, \&
  {Roth}}]{romero2018}
{Romero}, G.~E., {M{\"u}ller}, A.~L., \& {Roth}, M. 2018, \aap, 616, A57

\bibitem[{{Romero} {et~al.}(2010){Romero}, {Vieyro}, \& {Vila}}]{romero2010}
{Romero}, G.~E., {Vieyro}, F.~L., \& {Vila}, G.~S. 2010, \aap, 519, A109

\bibitem[{Ruiz-Lapuente {et~al.}(2016)Ruiz-Lapuente, The, Hartmann, Ajello,
  Canal, Röpke, Ohlmann, \& Hillebrandt}]{Ruiz_Lapuente_2016}
Ruiz-Lapuente, P., The, L.-S., Hartmann, D.~H., {et~al.} 2016, \apj, 820, 142

\bibitem[{{Rybicki} \& {Lightman}(1979)}]{rybicki1979}
{Rybicki}, G.~B. \& {Lightman}, A.~P. 1979, {Radiative processes in
  astrophysics}

\bibitem[{{Shakura} \& {Sunyaev}(1973)}]{shakura1973}
{Shakura}, N.~I. \& {Sunyaev}, R.~A. 1973, \aap, 500, 33

\bibitem[{{Sharma} {et~al.}(2007){Sharma}, {Quataert}, \& {Stone}}]{sharma2007}
{Sharma}, P., {Quataert}, E., \& {Stone}, J.~M. 2007, \apj, 671, 1696

\bibitem[{{Stawarz} \& {Petrosian}(2008)}]{stawarz2008}
{Stawarz}, {\L}. \& {Petrosian}, V. 2008, \apj, 681, 1725

\bibitem[{{Stecker} {et~al.}(1999){Stecker}, {Salamon}, \&
  {Done}}]{stecker1999}
{Stecker}, F.~W., {Salamon}, M.~H., \& {Done}, C. 1999, [arXiv:9912106], astro

\bibitem[{{Stepney} \& {Guilbert}(1983)}]{stepney1983}
{Stepney}, S. \& {Guilbert}, P.~W. 1983, \mnras, 204, 1269

\bibitem[{{Stone} {et~al.}(1999){Stone}, {Pringle}, \& {Begelman}}]{stone1999}
{Stone}, J.~M., {Pringle}, J.~E., \& {Begelman}, M.~C. 1999, \mnras, 310, 1002

\bibitem[{{Svensson}(1982)}]{svensson1982}
{Svensson}, R. 1982, \apj, 258, 335

\bibitem[{{Tchekhovskoy} {et~al.}(2011){Tchekhovskoy}, {Narayan}, \&
  {McKinney}}]{tchekhovskoy2011}
{Tchekhovskoy}, A., {Narayan}, R., \& {McKinney}, J.~C. 2011, \mnras, 418, L79

\bibitem[{{The Fermi-LAT collaboration}(2019)}]{fermi2019}
{The Fermi-LAT collaboration}. 2019, arXiv e-prints, arXiv:1905.10771

\bibitem[{{Toma} \& {Takahara}(2012)}]{toma2012}
{Toma}, K. \& {Takahara}, F. 2012, \apj, 754, 148

\bibitem[{{Tomsick} {et~al.}(2019){Tomsick}, {Zoglauer}, {Sleator}, {Lazar},
  {Beechert}, {Boggs}, {Roberts}, {Siegert}, {Lowell}, {Wulf}, {Grove},
  {Phlips}, {Brand t}, {Smale}, {Kierans}, {Burns}, {Hartmann}, {Leising},
  {Ajello}, {Fryer}, {Amman}, {Chang}, {Jean}, \& {von Ballmoos}}]{cosi2019}
{Tomsick}, J., {Zoglauer}, A., {Sleator}, C., {et~al.} 2019, in Bulletin of the
  American Astronomical Society, Vol.~51, 98

\bibitem[{{van Santen} \& {IceCube-Gen2 Collaboration}(2017)}]{vansanten2017}
{van Santen}, J. \& {IceCube-Gen2 Collaboration}. 2017, in International Cosmic
  Ray Conference, Vol. 301, 35th International Cosmic Ray Conference
  (ICRC2017), 991

\bibitem[{{Veledina} {et~al.}(2011){Veledina}, {Vurm}, \&
  {Poutanen}}]{veledina2011}
{Veledina}, A., {Vurm}, I., \& {Poutanen}, J. 2011, \mnras, 414, 3330

\bibitem[{{V{\'e}ron-Cetty} \& {V{\'e}ron}(2006)}]{veron2006}
{V{\'e}ron-Cetty}, M.~P. \& {V{\'e}ron}, P. 2006, \aap, 455, 773

\bibitem[{{Vieyro} \& {Romero}(2012)}]{vieyro2012}
{Vieyro}, F.~L. \& {Romero}, G.~E. 2012, \aap, 542, A7

\bibitem[{{Vila} {et~al.}(2014){Vila}, {Vieyro}, \& {Romero}}]{vila2014}
{Vila}, G.~S., {Vieyro}, F.~L., \& {Romero}, G.~E. 2014, in International
  Journal of Modern Physics Conference Series, Vol.~28, International Journal
  of Modern Physics Conference Series, 1460191

\bibitem[{{Vurm} \& {Poutanen}(2009)}]{vurm2009}
{Vurm}, I. \& {Poutanen}, J. 2009, \apj, 698, 293

\bibitem[{{Werner} {et~al.}(2018){Werner}, {Uzdensky}, {Begelman}, {Cerutti},
  \& {Nalewajko}}]{werner2018}
{Werner}, G.~R., {Uzdensky}, D.~A., {Begelman}, M.~C., {Cerutti}, B., \&
  {Nalewajko}, K. 2018, \mnras, 473, 4840

\bibitem[{{Willmer} {et~al.}(1991){Willmer}, {Focardi}, {Chan}, {Pellegrini},
  \& {da Costa}}]{willmer1991}
{Willmer}, C.~N.~A., {Focardi}, P., {Chan}, R., {Pellegrini}, P.~S., \& {da
  Costa}, N.~L. 1991, \aj, 101, 57

\bibitem[{{Wojaczy{\'n}ski} \& {Nied{\'z}wiecki}(2017)}]{wojaczynski2017}
{Wojaczy{\'n}ski}, R. \& {Nied{\'z}wiecki}, A. 2017, \apj, 849, 97

\bibitem[{{Wojaczy{\'n}ski} {et~al.}(2015){Wojaczy{\'n}ski}, {Nied{\'z}wiecki},
  {Xie}, \& {Szanecki}}]{wojaczynski2015}
{Wojaczy{\'n}ski}, R., {Nied{\'z}wiecki}, A., {Xie}, F.-G., \& {Szanecki}, M.
  2015, \aap, 584, A20

\bibitem[{{Wolstencroft} {et~al.}(1995){Wolstencroft}, {Done}, {Scarrott}, \&
  {Scarrott}}]{Wolstencroft1995}
{Wolstencroft}, R.~D., {Done}, C.~J., {Scarrott}, S.~M., \& {Scarrott},
  R.~M.~J. 1995, \mnras, 276, 460

\bibitem[{{Yarza} {et~al.}(2020){Yarza}, {Wong}, {Ryan}, \&
  {Gammie}}]{yarza2020}
{Yarza}, R., {Wong}, G.~N., {Ryan}, B.~R., \& {Gammie}, C.~F. 2020, \apj, 898,
  50

\bibitem[{{Yu} {et~al.}(2011){Yu}, {Yuan}, \& {Ho}}]{yu2011}
{Yu}, Z., {Yuan}, F., \& {Ho}, L.~C. 2011, \apj, 726, 87

\bibitem[{{Yuan} {et~al.}(2012){Yuan}, {Bu}, \& {Wu}}]{yuan2012}
{Yuan}, F., {Bu}, D., \& {Wu}, M. 2012, \apj, 761, 130

\bibitem[{{Yuan} \& {Narayan}(2014)}]{yuan2014}
{Yuan}, F. \& {Narayan}, R. 2014, \araa, 52, 529

\bibitem[{{Yuan} {et~al.}(2000){Yuan}, {Peng}, {Lu}, \& {Wang}}]{yuan2000}
{Yuan}, F., {Peng}, Q., {Lu}, J.-f., \& {Wang}, J. 2000, \apj, 537, 236

\bibitem[{{Yuan} {et~al.}(2003){Yuan}, {Quataert}, \& {Narayan}}]{yuan2003}
{Yuan}, F., {Quataert}, E., \& {Narayan}, R. 2003, \apj, 598, 301

\bibitem[{{Yuan} {et~al.}(2004){Yuan}, {Quataert}, \& {Narayan}}]{yuan2004a}
{Yuan}, F., {Quataert}, E., \& {Narayan}, R. 2004, \apj, 606, 894

\bibitem[{{Yuan} {et~al.}(2006){Yuan}, {Shen}, \& {Huang}}]{yuan2006}
{Yuan}, F., {Shen}, Z.-Q., \& {Huang}, L. 2006, \apjl, 642, L45

\bibitem[{{Yuan} \& {Zdziarski}(2004)}]{yuan2004b}
{Yuan}, F. \& {Zdziarski}, A.~A. 2004, \mnras, 354, 953

\bibitem[{{Zhdankin} {et~al.}(2019){Zhdankin}, {Uzdensky}, {Werner}, \&
  {Begelman}}]{zhdankin2019}
{Zhdankin}, V., {Uzdensky}, D.~A., {Werner}, G.~R., \& {Begelman}, M.~C. 2019,
  \prl, 122, 055101

\end{thebibliography}
%
%
\begin{appendix} 
\section{Accretion rate parametrization}
\label{ap:acc_rates}

The phenomenological function that parametrize the smooth transition between the SSD and the RIAF is
\begin{equation}
    f(R) = \left\{ 
    \begin{array}{lll}
    0 & {\rm if\ } R \le R_{\rm tr} \\
    \frac{1-\left(R_{\rm tr}/R\right)^b}{1-\left(R_{\rm tr}/R_{\rm out}\right)^b} & {\rm if\ } R_{\rm tr}<R\le R_{\rm out} \\
    1 & {\rm if\ } R_{\rm out}<R.
    \end{array}
    \right..
    \label{eq:f(R)}
\end{equation}
The normalized mass loss rate via winds $w(R)$ is
\begin{equation}
	w(R)= \dot{M}_{\rm out}^{-1}  \int_{R_{\rm out}}^R dR'~\frac{d \dot{M}}{d R'} \Bigg |_{\rm winds} = -s \int_{R_{\rm out}}^R dR' \frac{g(R')}{R'},
\end{equation}
where we have taken
\begin{equation}
	\frac{d \dot{M}}{d R} = s\frac{\dot{M}_{\rm c}(R)}{R}.
\end{equation}
Differentiating the relation $f(R)+g(R)=1-w(R)$, we obtain a first-order linear ordinary differential equation for $g(R)$:
\begin{equation}
	f'(R)+g'(R)= s \frac{g(R)}{R},
\end{equation}
whose solution is
\begin{equation}
\label{eq:g(R)}
	g(R) = \left( R/R_{\rm out} \right)^s - f(R) + s \int_R^{R_{\rm out}} dR' \left( \frac{R}{R'} \right)^s \frac{f(R')}{R'}.
\end{equation}
Inserting Eq. \ref{eq:f(R)} into Eq. \ref{eq:g(R)}, an analytical solution for $g(R)$ is easily obtained. This solution satisfy that $g(R)=0$ for $R>R_{\rm out}$ and $g(R) \propto (R/R_{\rm tr})^s$ for $R<R_{\rm tr}$.

\begin{figure}[h]
    \label{fig:accrates}
    \centering
    \includegraphics[width=0.9\linewidth]{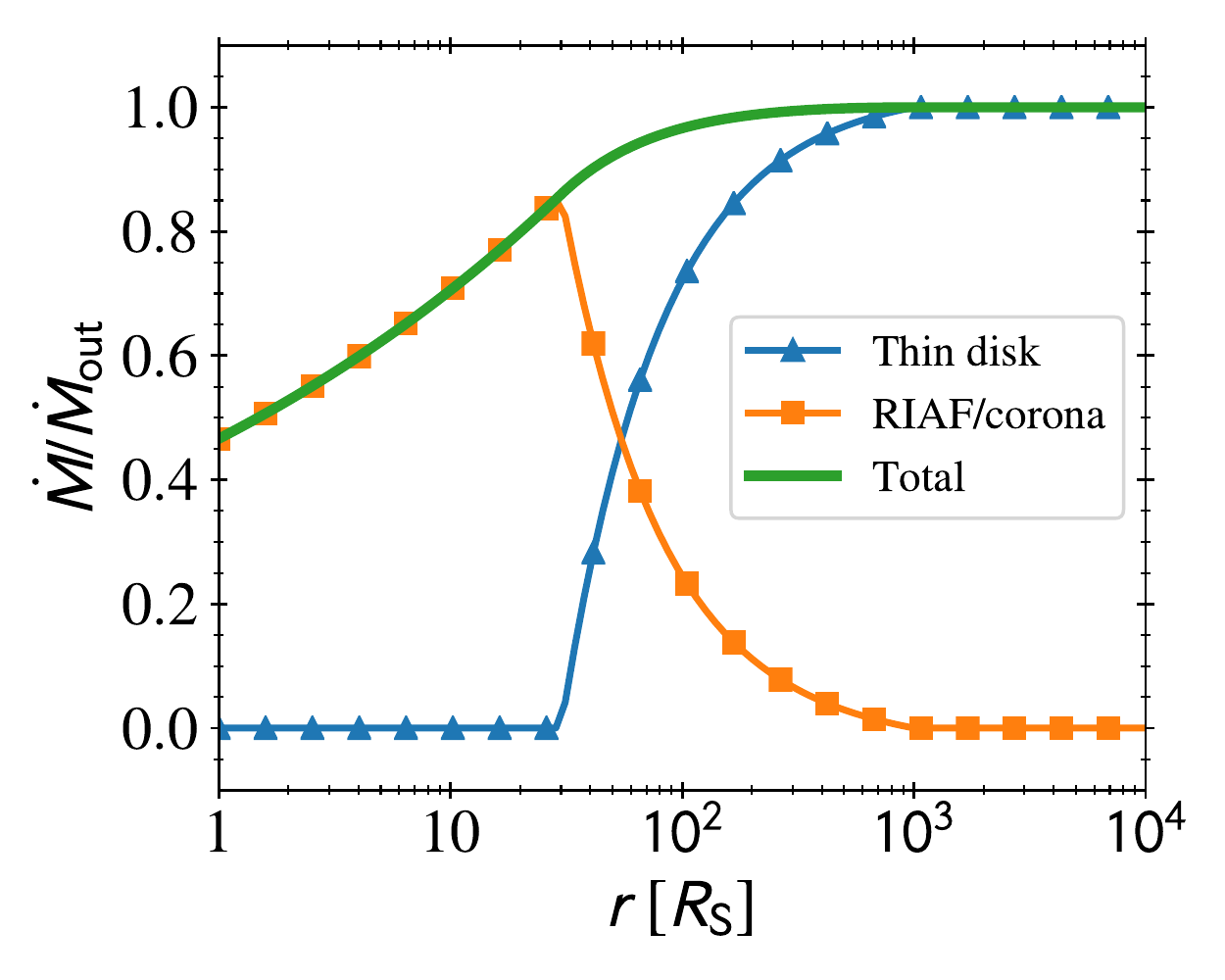}
    \caption{Accretion rates involved in the model for $R_{\rm tr}=30$, $R_{\rm out}=10^3$, $s=0.2$, and $\alpha=2$}
\end{figure}

Figure \ref{fig:accrates} shows the accretion rates as a function of the radius for $s=0.2$, $\alpha=2$, $R_{\rm tr}=30$, and $R_{\rm out}=10^3$.

\end{appendix}

\end{document}